# Exploring the Electronic Nature of Spinel Oxides: A Review of Their Electron Interactions and Prospects

Jagadis Prasad Nayak†, Raju Saini, Sourav Dash, and Gopi Nath Daptary*

Department of Physics and Astronomy, National Institute of Technology, Rourkela 769008, India

†jagadisnayak07@gmail.com;

*daptarygn@nitrkl.ac.in

**Abstract**

This review discusses the multifaceted electronic properties of spinel oxides with a particular focus on Lithium Vanadate ($LiV_2O_4$), Lithium Titanate ($LiTi_2O_4$), and Magnesium Titanate ($MgTi_2O_4$). We selected $LiTi_2O_4$, $LiV_2O_4$, and $MgTi_2O_4$ because they serve as quintessential examples of spinel oxides' diverse and intriguing electronic phenomena. $LiV_2O_4$ heavy fermion behaviour challenges traditional theories in $d$-electron systems, $LiTi_2O_4$ being the first oxide superconductor provides critical insights into unconventional superconductivity driven by strong electron-phonon interactions, and $MgTi_2O_4$ pronounced orbital ordering and metal-insulator transition offers a clear model for exploring electron-lattice coupling. This shows how the inherent structural versatility of the spinel lattice, characterised by its cubic close-packed oxygen network and variable cation distributions, enables a rich interplay of electron-electron correlations, electron-lattice coupling, and orbital degrees of freedom. In $LiV_2O_4$, the combination of mixed-valence vanadium ions and a geometrically frustrated pyrochlore lattice gives rise to heavy fermion behaviour, whereas $LiTi_2O_4$ exhibits unconventional superconductivity driven by a high density of states at the Fermi level and strong electron-phonon interactions. $MgTi_2O_4$ undergoes a pronounced metal-insulator transition, where orbital ordering triggers a Peierls-like distortion that stabilises a low-temperature insulating state through Ti-Ti dimerization. How the composition of these compounds affects their properties, based on both theoretical research and experimental findings, is illustrated by this review. It illustrates the promise of Spinels in practical technologies such as energy storage, electrocatalysis, and high-temperature lubrication.

## Introduction

Spinel oxides play a central position in condensed matter physics and materials science due to their extraordinary capability of connecting interesting electronic behaviours with structural versatility [1]. These compounds, characterised by their general formula $AB_2O_4$, are suitable for electron correlation effects, starting from heavy fermion behaviour and superconductivity to orbital ordering [2-4]. Within this diverse family, materials such as $LiV_2O_4$ (LVO), $LiTi_2O_4$ (LTO), and $MgTi_2O_4$ (MTO) have shown significant attention owing to their unique physical properties, which stem from the fundamental interactions between their electronic, magnetic, and structural degrees of freedom.

The role of electron-electron correlations, electron-lattice coupling, and orbital degrees of freedom collectively governs the emergent phenomena observed in these systems. LVO, for instance, is celebrated as the first oxide that exhibits heavy fermion behaviour, a property previously associated only with intermetallic compounds [2]. This research extends the limits of traditional comprehension through illustrating the way robust interactions between d-electrons conduct themselves in a geometrically frustrated lattice, a situation which creates rich and surprising physical phenomena. In this same context, LTO is unique for being the first known oxide to be superconductive. Its unorthodox character is strongly associated with an unusually elevated density of electronic states at the Fermi level, together with complex electron-phonon dynamics. In the case of MTO, on the other hand, interaction between lattice vibrations and electronic orbitals induces orbital ordering. This latter process, in its turn, induces structural phase transitions that profoundly affect the electronic transport properties of the material.

The spinel structure is central to controlling these electronic properties by offering a highly symmetric but accommodating structure that can accommodate a wide range of cation environments and valence states. This geometry typically results in magnetic frustration, which interferes with ordered magnetic behaviour and opens the door for the development of new states [5]. The presence of mixed-valence cations within these systems enhances electron-electron correlations, providing fertile ground for the development of complex phenomena like quantum criticality, non-Fermi liquid behaviour, and collective excitations that are beyond conventional description [6].

In addition to their basic scientific significance, spinel oxides have also become central materials for advanced technological innovation, with future-looking functionalities applied to an array of advanced applications [7]. Because of their ability to adapt their transport and magnetic behaviours while remaining thermodynamically stable, these materials hold great promise for use in energy storage, spintronic devices, and emerging quantum technologies. Exploring their electronic interactions not only deepens our understanding of how electrons behave in strongly correlated systems but also helps pave the way for developing the next generation of smart, high-performance materials [8, 9].

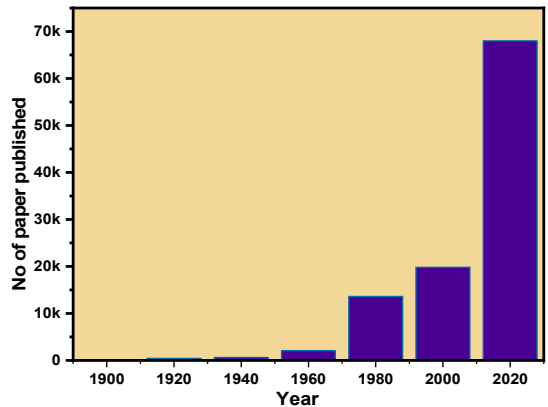

Figure 1. Recent research advancements on spinel oxides (compiled from published literature). Google Scholar. (2025, April 10). Search results for "spinel oxide" covering the years 1900 to 2025. Retrieved from https://scholar.google.com/scholar?q=spinel+oxide

As depicted in Figure 1, the publication record on spinel oxides has evolved significantly over the past century. Research activity in this field began modestly between 1900 and 1950, with only a few publications emerging during that early phase. As the decades progressed, particularly from the mid-twentieth century onward, the volume of literature steadily grew. A noticeable increase in the number of publications occurred in the 1990s, which increased till 2025. This upward trend not only reflects the expansion of research tools and methods but also points to the growing significance of spinel oxides in addressing modern technological needs, especially in areas like energy conversion and storage. It focuses on the electronic characteristics and behaviours of $LiV_2O_4$, $LiTi_2O_4$, and $MgTi_2O_4$, exploring how their structural features influence electron interactions and give rise to intriguing physical phenomena. In addition, it examines the current research landscape and identifies future directions that could bridge the gap between fundamental science and real-world applications. By integrating theoretical insights with experimental findings, this study aims to serve as a comprehensive resource for researchers seeking to uncover the intricate physics of spinel oxides and their potential in transformative technologies.

The roadmap of spinels presents the chronological development of spinel-based materials, highlighting key scientific discoveries and technological advancements as shown by Figure 2. Initially, spinel was distinguished from ruby in the early 1800s, marking the beginning of its independent study. The first spinel oxide was formed in $19^{th}$ century [10]. This milestone opened the door to controlled synthesis of materials, greatly influencing both the gemstone and ceramic industries. Then, in 1930, a key breakthrough came with the use of X-ray diffraction (XRD), which confirmed the cubic crystal structure of spinels and laid the foundation for understanding their crystallographic and electronic properties. In 1950, their magnetic properties had become a research subject, mainly for transformer cores and electronic components. The 1960s saw the rise of lithium-based spinels as promising cathode materials, leading to their application in lithium-ion batteries. This continues to play a decisive role in modern energy-storage technologies. A major milestone came in 1970 with the discovery of superconductivity in $LiTi_2O_4$, a rare phenomenon in oxide materials that sparked growing interest in spinel oxides as promising quantum materials. This breakthrough led to extensive research into their electronic correlations and unconventional superconducting behaviours. By the early 2000s, scientists began engineering nanostructured spinels for advanced uses in electronics, optics, and energy storage, taking advantage of their stability and adjustable properties. More recently, around 2020, research has pushed these limits even further, focusing on spinel-based multiferroics and quantum materials, thanks to new synthesis methods and innovative doping techniques. These developments underscore the versatility of spinels, not only in energy applications but also in next-generation computing and multifunctional materials research.

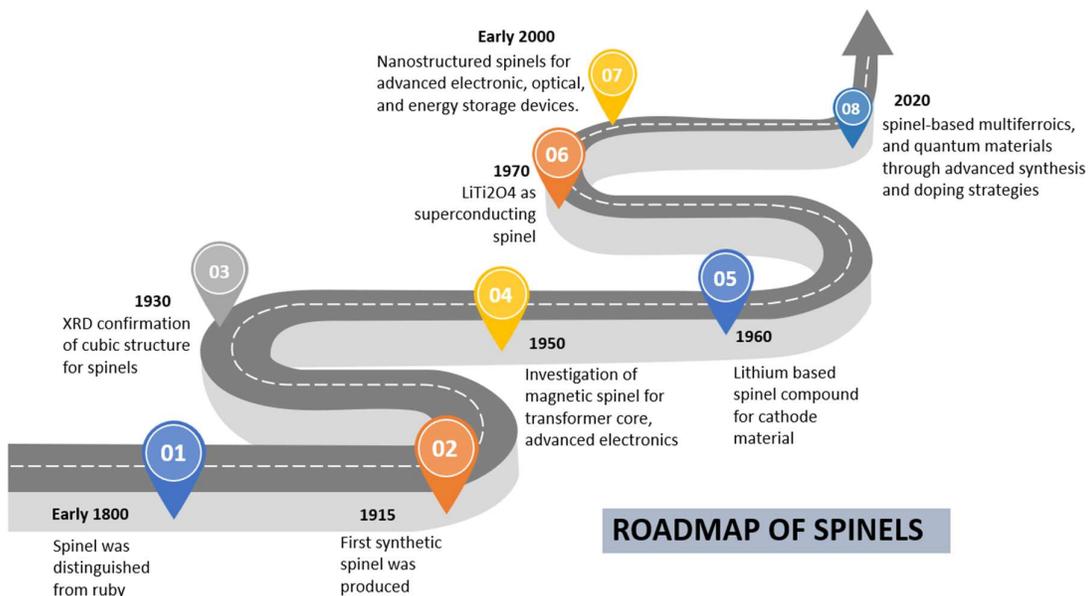

Figure 2 - A historical roadmap of spinel oxides, highlighting the major milestones, discoveries, and technological progress documented in the literature over time.

## 1.1 The Spinel Oxide Structure: A General Perspective

Spinel oxides crystallise in a cubic close-packed arrangement of oxygen anions, with metal cations distributed among the tetrahedral and octahedral interstitial sites. The general formula for spinel oxides is $AB_2O_4$, where the A-site cations occupy one-eighth of the tetrahedral voids. In contrast, the B-site cations fill half the available octahedral voids within the face-centred cubic (FCC) oxygen sublattice. [11]. This structure resulted in a highly ordered and symmetrical framework supporting various compositions and electronic configurations. The fundamental stability of the spinel lattice arises from the balance of the electrical interactions between the cations and anions, as well as the geometric constraints imposed by the oxygen arrangement.

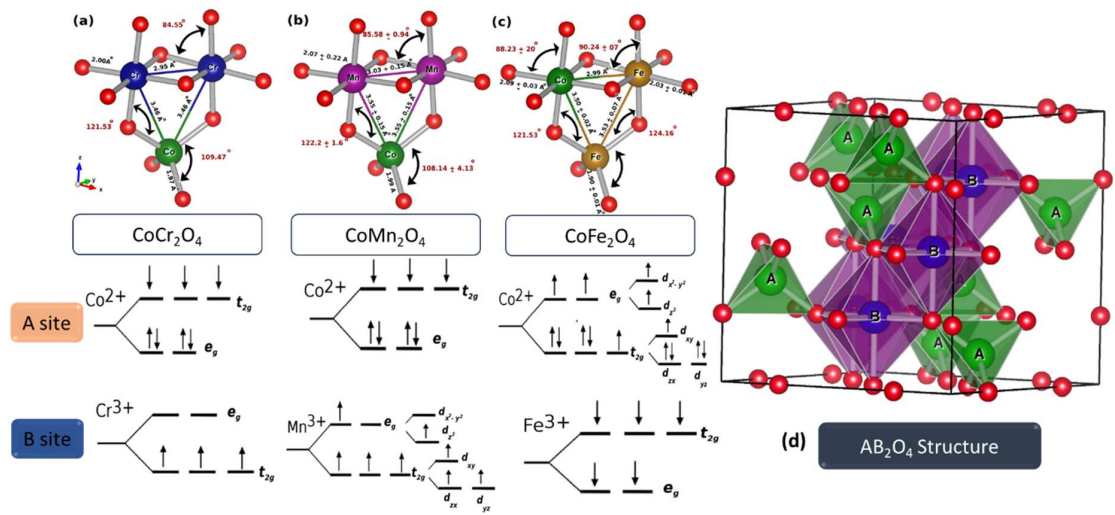

Figure 3. Illustration of the spinel $AB_2O_4$ structure and local cation environments for A = $Co^{2+}$ and B = [$Cr^{3+}$, $Mn^{3+}$, $Fe^{3+}$]. Panels (a)-(c) show the tetrahedral (A) and octahedral (B) coordination in $CoCr_2O_4$, $CoMn_2O_4$ and $CoFe_2O_4$, respectively, with bond lengths and angles. Below are schematic crystal-field diagrams for each cation (A- and B-site d-orbitals). On the right, the generic $AB_2O_4$ spinel framework is depicted, highlighting tetrahedral A sites (green) and octahedral B sites (purple) [12].

While $CoCr_2O_4$, $CoMn_2O_4$, and $CoFe_2O_4$ exemplify the classical spinel framework with well-defined, typical spinel structures and robust magnetic ordering driven by stable $AB_2O_4$ cation configurations, the focus of this review is on LTO, LVO, and MTO. Despite sharing the spinel structural idea, these Li- and Mg-based compounds exhibit unique electronic behaviours ranging from superconductivity in LTO to heavy fermion characteristics in LVO, stemming from their more complex and often partially inverted cation distributions, which lead to significant electronic relationships not present in the Co-based analogues. [13-15].

One of the standout characteristics of the spinel structure is its flexibility in accommodating different arrangements of cations, which gives rise to two main types: normal and inverse spinels. In a normal spinel, the divalent A-site cations settle into the tetrahedral sites, while the trivalent B-site cations occupy the octahedral positions [11]. In contrast, an inverse spinel has half of the trivalent B-site cations occupying the tetrahedral sites, while the remaining B-site cations, along with all the A-site cations, are found in the octahedral positions [11, 16]. The way cations are arranged in the spinel structure has a big impact on the material's electrical, magnetic, and catalytic properties. By carefully choosing which cations to include and controlling their distribution, these materials become highly adaptable for a wide range of technological uses. But the tunability of spinels goes beyond just cation ordering, partial substitutions, and doping at both the A and B sites, offering even more ways to tailor their physical and chemical traits. This flexibility has made spinel oxides a hot topic in research, with applications spanning battery electrodes, magnetic materials, catalysts, and transparent conductors. By tweaking their composition, scientists can fine-tune properties like electrical conductivity, magnetoresistance, and catalytic activity. Plus, their inherent stability under harsh conditions makes spinel oxides especially valuable for energy storage, electronics, and environmental technologies.

1. **Structural Characteristics of LVO, LTO, and MTO**

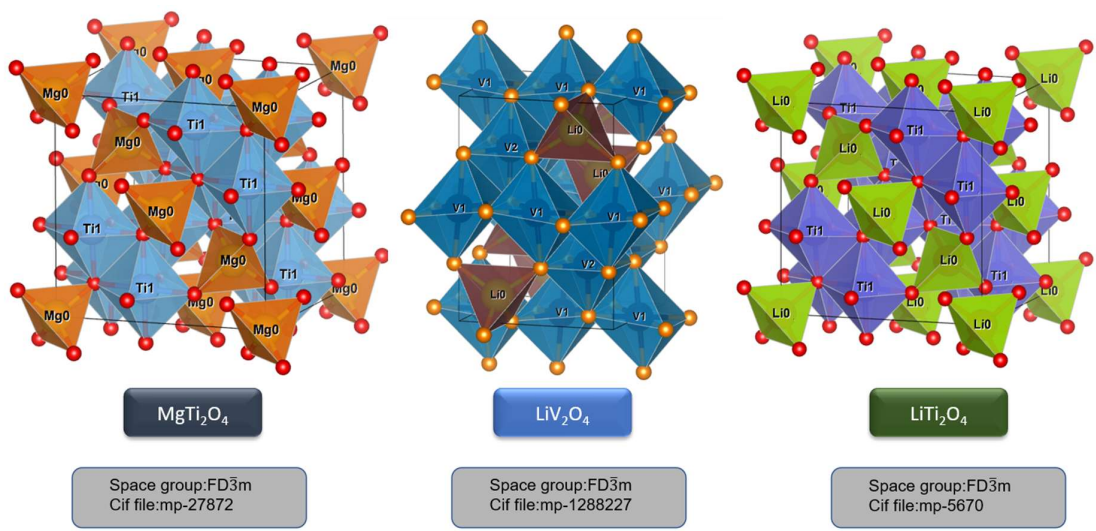

$MgTi_2O_4$ — Space group: $FD\bar{3}m$, Cif file: mp-27872

$LiV_2O_4$ — Space group: $FD\bar{3}m$, Cif file: mp-1288227

$LiTi_2O_4$ — Space group: $FD\bar{3}m$, Cif file: mp-5670

Figure 4. Structures of three spinel compounds MgTi$_2$O$_4$, LiV$_2$O$_4$, and LiTi$_2$O$_4$, all crystallising in space group Fd-3m. The orange polyhedra highlight the tetrahedral (A) sites, and the blue polyhedra highlight the octahedral (B) sites. Each compound's CIF file (from the Materials Project) is indicated below the structure.

**2.1 LVO:** is a unique compound that crystallises in the spinel structure, where lithium ions occupy the tetrahedral A sites and vanadium ions reside in the octahedral B sites. In this configuration, the vanadium ions form a pyrochlore lattice, a network of corner-sharing tetrahedra known for its inherent geometric frustration. This frustration arises because the triangular arrangement of magnetic ions prevents the system from easily settling into a simple magnetic ground state, leading to a highly degenerate ground state with competing interactions [17-19]. LVO is a special material with heavy fermion behaviour below 10 K, yet it is a *d*-electron metal. This makes it an unusual material, because d orbitals have a larger spatial extent than *f* orbitals. Additionally, their hybridisation with conduction electrons is also stronger. The structure of LVO is FCC (face-centred cubic), which belongs to the Fd$\overline{3}$m space group. The crystal structure is shown in Figure 4. It has a 'normal' spinel structure, where vanadium atoms are surrounded by six oxygen atoms, forming a slightly distorted octahedron.

In LVO, the formal oxidation state of vanadium is non-integer, approximately V$^{3.5+}$ [20, 21], resulting in an effective $d^{1.5}$ electronic configuration with all V ions crystallographically equivalent. In the AB$_2$O$_4$ spinel structure, lithium occupies the tetrahedral A sites, while the vanadium ions form a highly frustrated pyrochlore lattice at the octahedral B sites. A slight trigonal distortion in the VO$_6$ octahedra causes the t$_{2g}$ orbitals to split into a more localised a$_{1g}$ orbital and a more itinerant e′$_g$ orbital. The interaction between these itinerant and localised electrons, often described as Kondo coupling, is invoked to explain the heavy fermion behaviour observed in LVO, ultimately accounting for its inherently metallic nature despite strong electronic correlations. [22, 23].

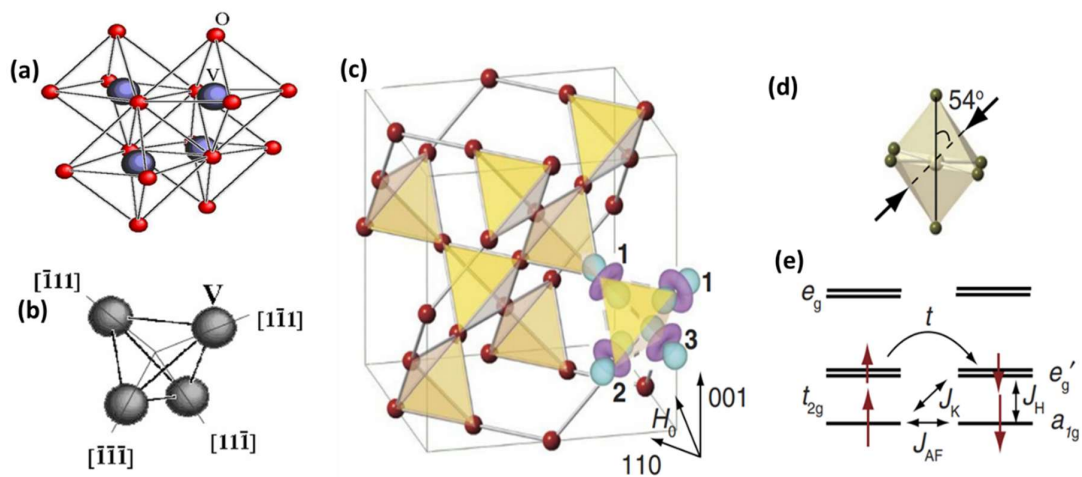

Figure 5 (a) LVO structure [24]. (b) Tetrahedron formed from four V atoms in the spinel unit cell and the trigonal axes for each V atom [24]. (c) The corner-sharing V tetrahedral structure leads to geometric frustration, inhibiting magnetic ordering. In the cubic lattice with local trigonal distortion, three vanadium sites (1, 2, and 3) with a$_{1g}$-like orbitals are inequivalent when the magnetic field $H_0$ rotates from [001] to [110]. (d) A trigonally distorted $VO_6$ ligand field splits the fivefold-degenerate $3d$ orbitals into $e_g$ and $t_{2g}$, where $t_{2g}$ further separates into a lower $a_{1g}$ and an upper $e'_g$ level, accommodating 1.5 electrons. For sites 2 and 3, the trigonal axis is tilted 54° from the [001] crystal axis. (e) In a Kondo lattice, localised $a_{1g}$ and itinerant $e'_g$ electrons interact through on-site ferromagnetic Hund's coupling ($J_H$), off-diagonal Kondo coupling ($J_K$), and off-site antiferromagnetic kinetic exchange ($J_{AF}$) [22].

A prominent feature of LVO is the mixed valence state of the vanadium ions, specifically $V^{3+}$ and $V^{4+}$. This mixed valence leads to strong electron-electron relationships because electrons can fluctuate between valence states. Such fluctuations are a key factor in the emergence of heavy fermion behaviour, where the effective mass of the conduction electrons is significantly improved. In LVO, this behaviour is observed below a characteristic temperature of approximately 20 K, where the material exhibits properties akin to those found in $f$-electron heavy fermion systems, despite being a $d$-electron system. [22]. The geometric frustration inherent in the pyrochlore lattice of LVO plays a crucial role in determining its electronic properties. This prevents the establishment of long-range magnetic order, even at very low temperatures, leading to a paramagnetic state down to temperatures as low as $80 \times 10^{-2}$ K[25]. This suppression of magnetic ordering enhances spin fluctuations, which, in conjunction with the mixed valence state, contribute to the heavy fermion behaviour observed in the material. The overall space group of this crystal is cubic, but the local point group symmetry is [23, 24] of the crystallographic position of the V ion is trigonal $D_{3d}$. The trigonal axes of each V atom in the unit cell are oriented towards the centre of the tetrahedron.

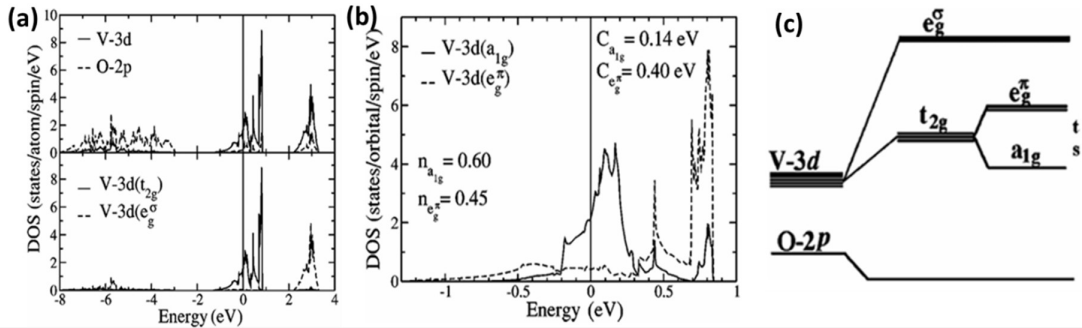

Figure 6 (a) Calculated total and partial density of states (DOS) for LVO, highlighting the contributions from V 3d and O 2p orbitals. (b) Detailed view of the V 3d-derived DOS near the Fermi level, illustrating the occupation ($n \approx 0.45$) and crystal-field splitting. (c) Schematic energy-level diagram for the V 3d states in an octahedral environment, showing the $t_{2g}$–$e_g^\sigma$ crystal-field splitting and the additional splitting into $a_{1g}$, $e_g^\pi$ orbitals [24].

From the Figure 6 above the occupation numbers and crystal-field splitting in panel (b) are determined from the projected density of states (PDOS) calculations, typically obtained using density functional theory (DFT). The DOS plot shows the contributions of different vanadium $3d$ orbitals ($a_{1g}$ and $e_g^\pi$) near the Fermi level, with the occupation numbers estimated by integrating the DOS up to the Fermi energy. The crystal-field splitting energies are determined from the energy separation of these orbitals, which arises due to the local octahedral coordination and additional distortions such as trigonal splitting. The schematic in panel (c) illustrates this crystal-field effect, showing the splitting of V 3d states into $t_{2g}$ and $e_g^\sigma$ levels, with further splitting within $t_{2g}$ into $a_{1g}$ and $e_g^\pi$ due to lower symmetry effects. The partial LDA density of states (DOS) for the $a_{1g}$ and $e_g^\pi$ orbitals. The $a_{1g}$ orbital has a bandwidth of 1.35 eV, nearly half of the $e_g^\pi$ bandwidth of 2.05 eV. Despite this, LDA calculations suggest that the electron population in all $t_{2g}$-derived orbitals is nearly identical, with 0.60 electrons in $a_{1g}$ and 0.45 in $e_g^\pi$ [24]. The spectral weight of the $a_{1g}$ orbital is concentrated around the Fermi level, between -0.2 eV and 0.3 eV, whereas the $e_g^\pi$ DOS is more evenly distributed and primarily ranges from 0.3 eV to 0.85 eV. The trigonal crystal field splitting, determined from the energy difference between the centres of gravity of the $a_{1g}$ and $e_g^\pi$ states, indicates that the $a_{1g}$ orbital is more favourable for electron localization. The effective mass, which reflects the impact of Coulomb interactions, can be inferred from the electronic specific heat coefficient ($\gamma$) at low temperatures [24]. The mass enhancement ratio, $\frac{m^*}{m_b} = \frac{\gamma}{\gamma_{LDA}}$, is a measure of electron correlation effects in materials like LVO. Here, $m^*$ represents the effective mass of quasiparticles, which accounts for electron-electron interactions and is typically extracted from experimental specific heat measurements. In contrast, $m_b$ is the band mass, obtained from band structure calculations, often using the local density approximation (LDA). The electronic specific heat coefficient $\gamma$ is determined experimentally, while $\gamma_{LDA}$ is its theoretical counterpart from LDA calculations. A significant enhancement in $m^*$ relative to $m_b$ suggests strong electron correlations, as seen in heavy fermion and other correlated electron systems. Using LDA-calculated $\gamma_{LDA}$, the mass enhancement ratio was found approximately as 25.8 for LVO, matching with previous studies [24]. This significant improvement in effective mass suggests that Coulomb interactions play a crucial role in LVO and must be considered to fully understand its electronic properties.

**2.2 LTO:** In LTO, which crystallizes in the normal spinel structure $AB_2O_4$, the Li cations occupy the tetrahedral A-sites, forming a corner-sharing tetrahedral network known as a pyrochlore lattice. Meanwhile, the Ti cations exist in the octahedral B-sites, where they are coordinated by six oxygen anions, forming $TiO_6$ octahedra. This structural arrangement leads to a complex connectivity between the oxygen polyhedra. A defining feature of the spinel structure is the way the tetrahedral $LiO_4$ and octahedral $TiO_6$ units share corners. The oxygen anions in the $TiO_6$ octahedra are also shared with the $LiO_4$ tetrahedra, creating a unique connectivity. This arrangement places specific constraints on the positions of the oxygen atoms; they need to satisfy the bonding requirements of both tetrahedral and octahedral sites. As a result, the oxygen ions shift slightly from their ideal octahedral positions to strike a balance between the bond lengths and angles in these two different environments.

This structural adjustment leads to a triangular distortion in the $TiO_6$ octahedra. Instead of maintaining a perfect octahedral symmetry, the Ti-O bond angles are slightly altered due to the influence of the Li sublattice. This distortion is particularly relevant in spinel oxides, as it affects electronic interactions, orbital overlap, and potentially plays a role in the superconducting behaviour of LTO by modifying the electron hopping pathways within the Ti sublattice.

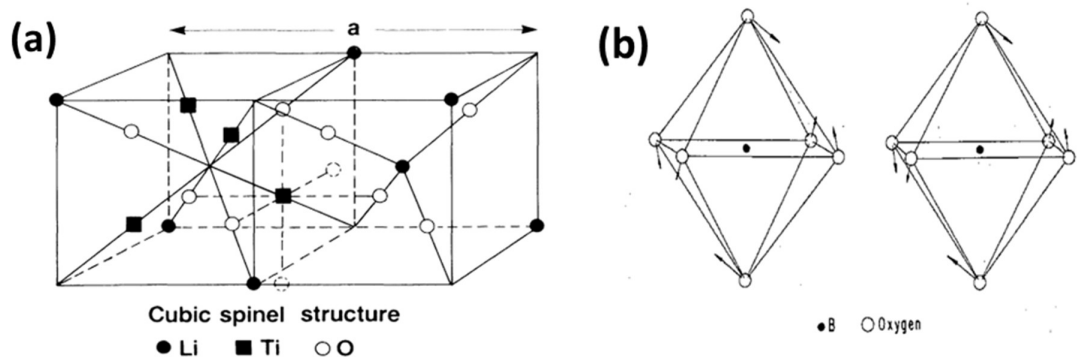

Figure 7 (a) The cubic spinel structure of LTO. The atoms shown here form the basis of a face-centred-cubic Bravais lattice with "a" being the lattice constant. (b) Distortion of the octahedron in Spinel surrounding B/Ti cation [26, 27].

LTO features titanium ions at the B-sites in various oxidation states. Its superconducting properties are deeply tied to the electronic density of states (DOS) at the Fermi level and the intricate interplay between its lattice and electronic subsystems. The simple cubic structure of this material provides a clean platform for studying the mechanisms of oxide superconductivity. Due to the low melting point of $Li_2CO_3$/LiOH precursor used, Lithium (Li) evaporates from the sample making it off-stoichiometric and difficult to synthesize in Polycrystalline form. Due to deficiency of Li in the sample many of its electronic and structural properties change. But it has been seen that the pure phase $LiTi_2O_4$ is metallic in nature and as the stoichiometry deviates it transforms into an insulating state showing Metal Insulator Transition [28].

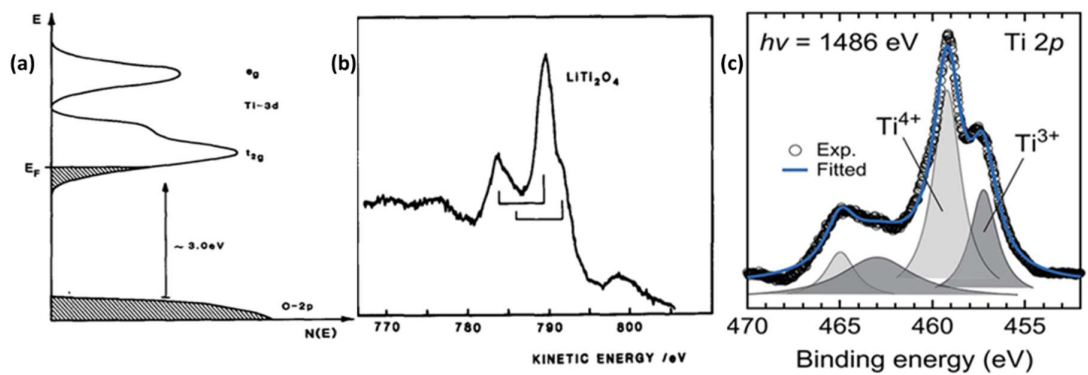

Figure 8. (a) Schematic density of states for spinel LTO, illustrating partially occupied Ti 3d states near the Fermi level and O 2p states at lower energy. (b) Representative photoemission spectrum showing the valence band and conduction band features, highlighting the contribution of Ti 3d electrons [29]. (c) High-resolution Ti 2p X-ray photoelectron spectrum ($hv$ = 1486 eV) deconvolved into $Ti^{3+}$ and $Ti^{4+}$ components, confirming the mixed-valence nature of LTO [30].

Figure 8 shows a comprehensive spectroscopic analysis of LTO's electronic structure. The left panel shows the valence band photoemission spectrum, where the dominant feature near the Fermi level is attributed to Ti 3d electrons, and additional peaks at lower binding energies arise from O 2p states. This atomic arrangement highlights the significant Ti-O hybridisation and its impact on the material's conductivity. The centre panel shows the conduction band region, where the gradual increase in spectral weight near the Fermi level indicates the available states for charge carriers, key to understanding how conduction happens and how superconductivity emerges. On the right, the high-resolution Ti 2p core-level spectrum has been carefully broken down into $Ti^{3+}$ and $Ti^{4+}$ components. The presence of this mixed-valence state, seen through the relative intensities and binding energies of the peaks, points to strong electron correlation effects. This mixed valence is a crucial factor behind the metal-insulator transition observed in $LiTi_2O_4$. Together, these results provide a detailed view of how electronic states, orbital hybridisation, and electron correlations interact in LTO, shedding light on the mechanisms driving its phase changes and superconducting properties.

Using X-ray Diffraction methods, it has been studied that Li occupies all of the tetrahedral sites and Ti occupies most of the octahedral sites, but still there is partial occupancy of Li in the octahedron, making the polycrystalline sample off-stoichiometric $Li_{1+x}Ti_{2-x}O_4$ [31]. The Ti ion's $3d$ orbitals are split into more stable $t_{2g}$ orbitals and less stable $e_g$ orbitals by the crystal field energy. From the band structure **(Fig. 8. (a))**, it was clearly seen that for LTO, the conduction band lies in the $t_{2g}$ orbital. The X-ray Photo Emission spectra of LTO **(Fig. 8. (b) & (c))** shows overlapping of two spin orbit doublets which indicates the presence of two distinct oxidation states of Transition metal ion, i.e., $Ti^{3+}$ ($d^1$) and $Ti^{4+}$ ($d^0$) in the system, with each Ti ion occupying $Ti^{3.5+}$ ($3d^{0.5}$) configuration.

**2.3 MTO:** MTO is a spinel oxide that shows a notable metal-insulator transition at approximately 260 K, accompanied by a structural change from cubic to tetragonal symmetry. This transition is primarily driven by the orbital ordering of the titanium ions at the octahedral B sites. In the high-temperature cubic phase, $Ti^{3+}$ ions possess a degenerate electronic configuration, leading to orbital degeneracy. As the temperature decreases, the system relieves this degeneracy through cooperative Jahn-Teller distortion, resulting in an ordered arrangement of orbitals. This ordering leads to the formation of Ti-Ti dimers, effectively pairing the Ti ions and causing a Peierls-like distortion that stabilizes the low-temperature insulating phase [32].

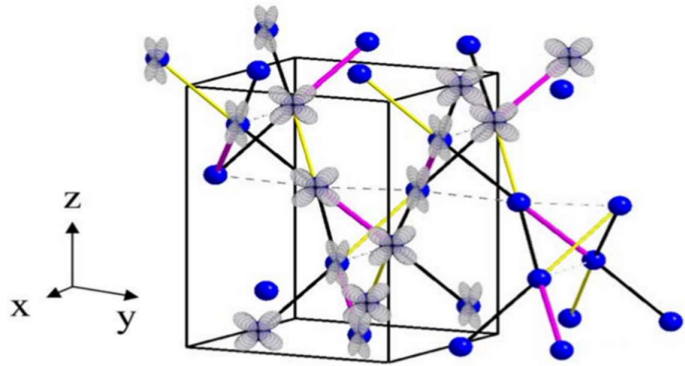

Figure 9. At low temperatures, MTO exhibits a specific pattern of orbital ordering. In this structure, B-site titanium ions (represented as blue dots) form chains through corner-sharing tetrahedra. Within these chains, bonds of varying lengths are observed: short (depicted in pink), intermediate (black), and long (yellow). The $d_{YZ}$ and $d_{ZX}$ orbitals align along these chains, specifically in four crystallographic directions: [0,1,1], [0,1, -1], [1,0,1], and [1,0, -1] [33].

The Peierls-like distortion in MTO involves dimerization of Ti ions along specific crystallographic directions, leading to periodic lattice distortion. This dimerization opens a gap at the Fermi level, transitioning the material from a metallic to an insulating state. The strong coupling between the lattice vibrations and orbital degrees of freedom is central to this process, as the lattice distortion is both a consequence of and a driving force for orbital ordering. This interplay between the electronic and lattice dynamics is a hallmark of the orbital-Peierls state observed in MTO [33].

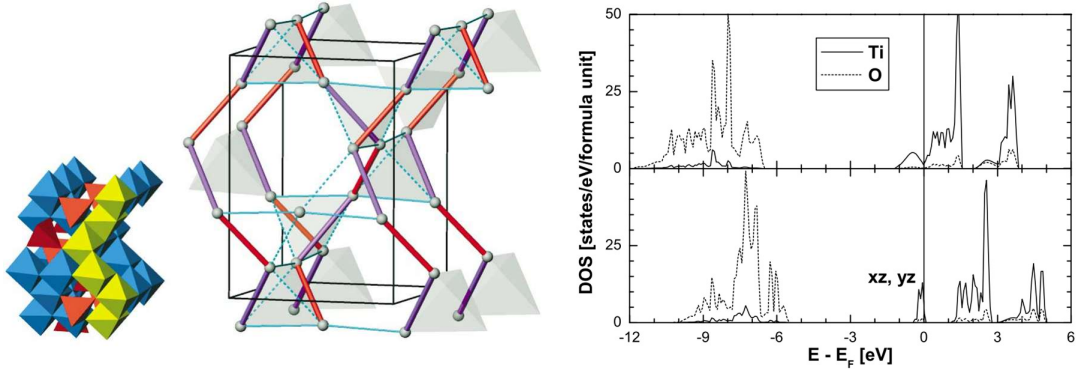

Figure 10. (a) In tetragonal MTO at 200 K, Ti–Ti bonds dimerize: red (shortest), purple (longest), and blue (intermediate, $i_1 \approx 3.0075$ Å, $i_2 \approx 3.0147$ Å). The inset shows a spinel fragment with a yellow-outlined "helix" [32]. (b) DOS for cubic (top) and tetragonal (bottom) MTO, with distinct line styles for Ti and O (Mg is negligible); the vertical line indicates the Fermi level [32].

As shown in Figure 10 (a), at 200 K, the Ti–Ti bond network in tetragonal $MgTi_2O_4$ shows clear variations in bond lengths, revealing a structural distortion. The dimerized Ti-Ti pairs (highlighted by red bonds) point to a Peierls-like distortion, with the longest bonds (purple) and intermediate bond lengths (around 3.0075 Å and 3.0147 Å) forming a non-uniform pattern. This distortion alters the connectivity between Ti tetrahedra and affects the material's electronic correlations. The inset emphasises a helical arrangement of Ti atoms, a hallmark of spinel structures, which could influence orbital ordering and charge transport. Understanding these changes is key to unravelling phenomena like metal-insulator transitions, superconductivity, and heavy fermion behaviour in spinel oxides [32].

The orbital ordering and associated Peierls distortion have significant implications for the electronic transport properties of MTO. Above the transition temperature, the material exhibited metallic conductivity owing to the delocalized nature of the Ti 3d electrons. Below the transition, the opening of the energy gap due to dimerisation leads to insulating behaviour, which is characterised by a sharp decrease in electrical conductivity. This transition is also accompanied by changes in magnetic properties, as the formation of spin-singlet pairs within the Ti-Ti dimers leads to a reduction in magnetic susceptibility [34].

2. **Electronic behaviour in LTO, LVO & MTO**

### 3.1 LTO: Electron Behaviour and Superconductivity

The superconducting properties of LTO are significantly influenced by its high density of states (DOS) at the Fermi level and the mixed-valence nature of Ti ions. These factors enhance electron-phonon coupling, which is crucial for superconductivity, as shown by various spectroscopic studies and isotope-effect measurements [35]. The DOS at the Fermi level is critical for decisive the electronic properties of materials. A high DOS indicates greater availability of electronic states for conduction, which enhances electron-phonon interactions [36, 37]. Spectroscopic studies have shown that the electron-phonon coupling in LTO is substantial, supporting the phononic mediation of superconductivity. This coupling is essential for the formation of Cooper pairs that are fundamental to the superconducting state.

In LiTi$_2$O$_4$, titanium exists in mixed-valence states (Ti$^{3+}$/Ti$^{4+}$), which allows electrons to delocalize. This delocalisation creates a metallic conduction pathway, boosting the material's electrical conductivity [38]. This mixed-valence state also gives rise to strong electronic correlations, which contribute to the unconventional superconducting properties of LiTi$_2$O$_4$. These correlations lead to unique electronic behaviours that aren't usually seen in conventional superconductors. While the high density of states and mixed-valence nature are key factors for superconductivity, it's important to remember that other elements, like the crystal structure and external conditions, can also have a significant impact on LTO's superconducting behaviour.

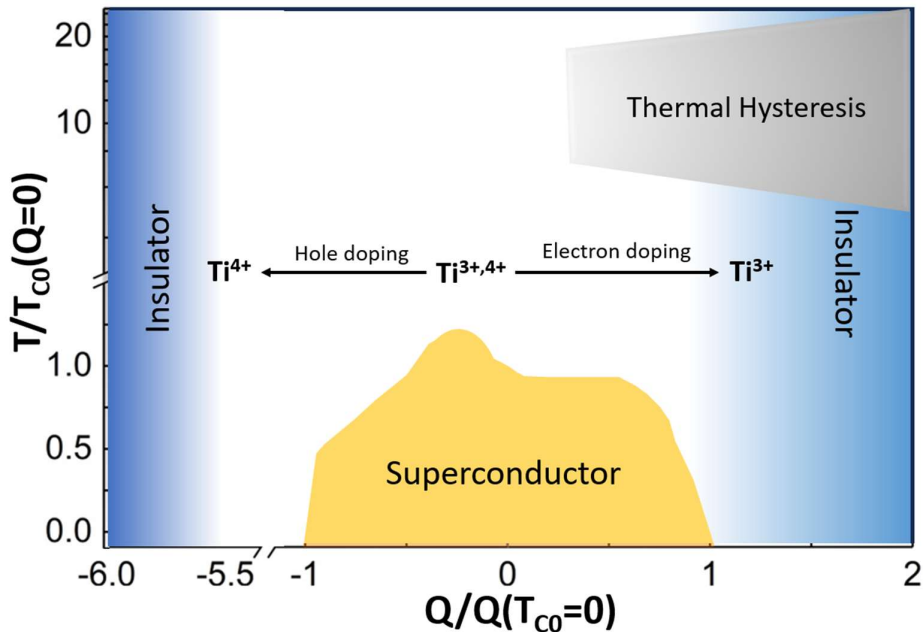

Figure 11. Phase diagram of LTO plotted as a function of normalized charge carrier doping $Q/Q$ (T$_{C0}$ = 0) and normalized temperature T/T$_{C0}$(Q = 0) [28].

In the Figure 11, Q denotes the effective charge carrier concentration, which is tuned via Ti valence (Ti$^{3+}$ ↔ Ti$^{4+}$) through electron or hole doping, and $T$ is the temperature. The yellow region marks the superconducting phase, bounded by insulating phases on both the electron-doped (Ti$^{3+}$) and hole-doped (Ti$^{4+}$) sides. The mixed-valence regime (Ti$^{3+}$/Ti$^{4+}$) supports superconductivity [28].

The electronic band structure calculations were performed on LTO using the all-electron full potential linear augmented plane wave plus local orbitals (FP-LAPW+LO) method. Figure 12 (a) displays electronic band structure for bulk LTO within the framework of density functional theory. With and without spin-orbit coupling the energy difference arises to be 19.4 meV per unit cell. The Density of states (DOS) was also calculated on non-magnetic LTO, which indicates the conduction electrons near the fermi level mainly come from the Ti 3d orbital partially hybridized with O 2p orbitals. By verifying the orbitals resolved energy band and project DOS (PDOS), the $3d_{x^2-y^2}$ and $3d_{xy}$ orbitals significantly contribute to the electronic conductions near the fermi level.

When the photoemission spectra of a material are studied, we are essentially looking at how electrons are distributed in different energy levels. In this case, the measurements show that there is about a 3.3 eV gap between the energy levels of electrons in the titanium (Ti) $3d$ orbitals and those in the oxygen (O) $2p$ orbitals. This gap represents the energy needed for electrons to move from the O 2p states (typically forming the valence band) up to the Ti 3d states (which contribute to the conduction band). In simpler terms, most of the electrons that are free to move and carry electrical current in the spinel structure are found in the Ti 3d states. This experimental finding is supported by theoretical calculations (first principles calculations) that map out the DOS of the material. The DOS analysis confirms that the Ti 3d states are the main contributors to the conduction properties, meaning these electrons are the primary charge carriers that help conduct electricity in the spinel.

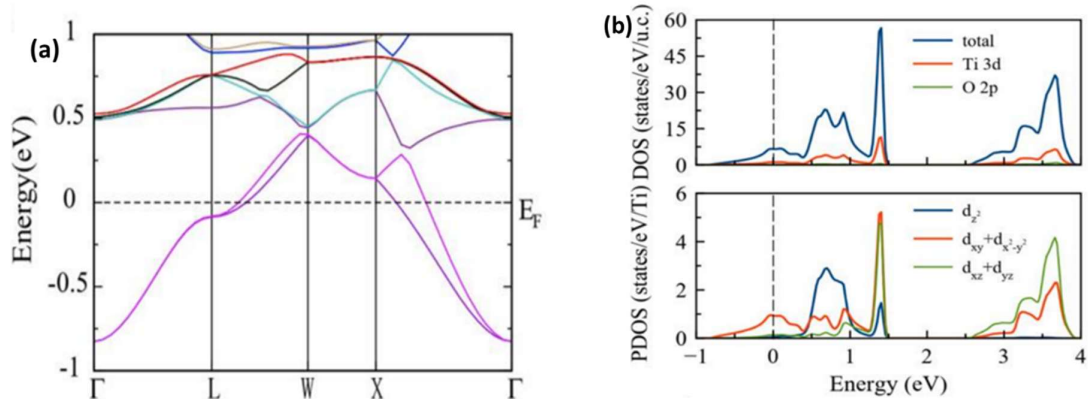

Figure 12. (a) Electronic band structure of LTO (b) DOS & Projected-DOS with Spin-orbit Coupling [39].

Electron-phonon coupling in LTO is particularly strong and plays a crucial role in its unconventional superconductivity. This interaction helps form the Cooper pairs necessary for the superconducting state. Although LTO shows a non-magnetic ground state when superconducting, subtle magnetic fluctuations suggest there are complex interactions between spin dynamics and superconductivity at play [40]. The electron-phonon coupling in $LiTi_2O_4$ is anisotropic, with certain crystallographic orientations exhibiting notably stronger interactions [41]. Raman studies revealed multiple active modes, indicating strong electron-phonon interactions, particularly anomalies at the magnetoresistance transition [42]. The unconventional d-wave-like pairing mechanism is supported by the observed fourfold symmetry in superconducting properties. The interplay between spin dynamics and superconductivity proposes that magnetic interactions could play a role in the pairing mechanism, despite the nonmagnetic nature of the ground state [41].

*Frustrations in LTO Spinel Oxide*

In some crystal lattices, accomplishing a configuration where all interactions are simultaneously satisfied is impossible, leading to a phenomenon known as frustration. Instead of a single, unique ground state, these frustrated systems exhibit multiple low-energy states, where energy non-minimization is shared across the system [43]. To illustrate this, consider a square lattice. In an antiferromagnetic system, spins align in an alternating pattern, resulting in a stable ground state. However, in a triangular lattice, if two spins are oriented antiparallel, the orientation of the third spin becomes ambiguous. Regardless of its alignment, the system cannot achieve a macroscopically stable state; instead, it exists in a mixed state with inherent dissatisfaction.

Frustrated systems often display metastability and hysteresis. Geometrical frustration can disrupt spin ordering, as observed in 2D Kagome lattices and 3D pyrochlore structures, which feature corner-sharing tetrahedra [44, 45]. These systems tend to exhibit a classical ground state with macroscopic degeneracy, known as cooperative paramagnetism, characterized by short-range spin correlations across all temperatures.

***Superconductivity feature:*** $LiTi_2O_4$ is the only oxide spinel to show super-conductivity and among the other superconducting spinels, it has a higher transition temperature in the range (11-13) K [46, 47]. From the resistivity and DC magnetization measurements for both polycrystalline and thin films, this system is a conventional s-wave BCS superconductor.

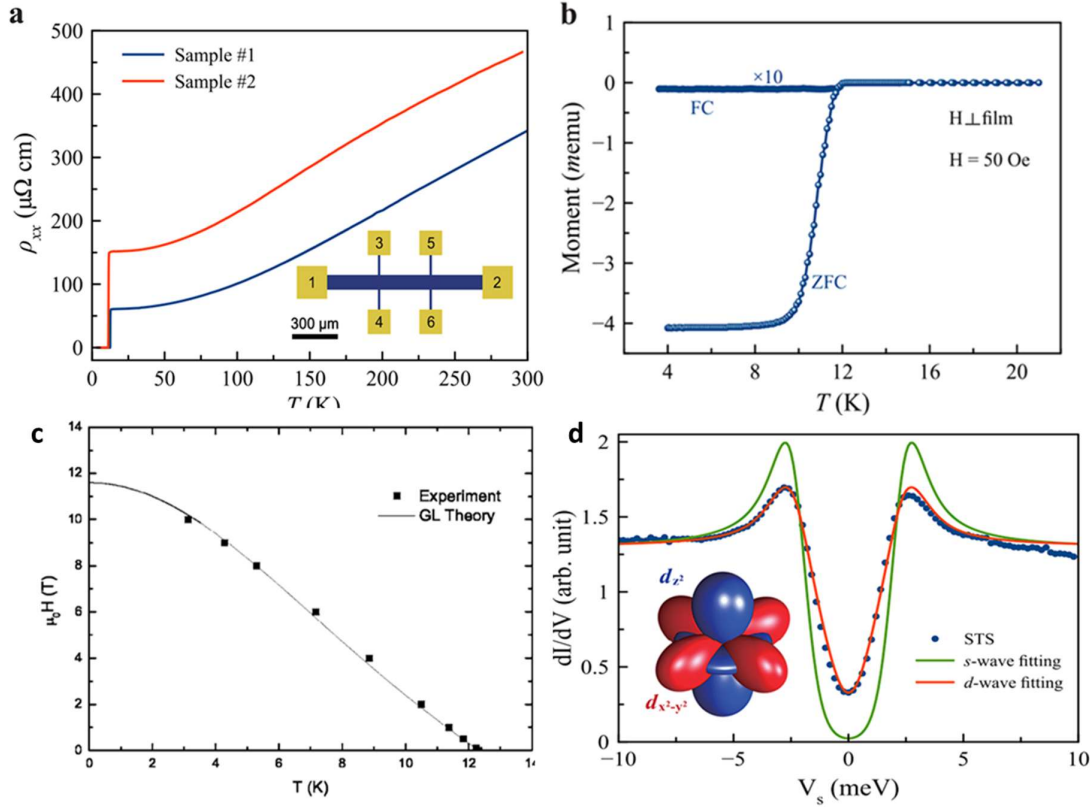

Figure 13. (a) Electrical resistivity of LTO as a function of temperature, (b) DC magnetization measurements in both zero-field cool (ZFC) and field cool (FC) modes. (c) $H_{c2}$-T phase diagram, solid line represents the best fitting from the Ginzberg-landau theory (d) High-resolution tunnelling conductance spectrum (dI/dV, $2\Delta_g$ = 4.8 meV) on the surface of $LiTi_2O_4$ at 4.2 K obtained using scanning tunnelling spectroscopy (STS) [39].

***Type-II superconductivity in LTO***

LTO is classified as a type-II superconductor because experimental measurements reveal a wide magnetic field range over which the superconducting state persists in the form of quantized vortices. In particular, low-temperature specific heat data show that the upper critical field $H_{C2}$ is approximately 11.7 T while the lower critical field $H_{C1}$ is around 26 mT [31]. This large disparity, along with a high Ginzburg–Landau parameter (k ≈ 29), indicates that magnetic flux penetrates the material as vortices rather than being completely expelled-a hallmark of type-II superconductivity. The data, which were well described by Werthamer-Helfand-Hohenberg (WHH) theory and are consistent with a fully gapped, moderate coupling BCS-like state, provide strong evidence for type-II behavior in LTO [48].

For confirming its type-II BCS superconductivity, the upper critical field, $H_{c2}$ vs T graph is determined from the ρ-T curves at different fields. From the Ginzberg-Landau theory the upper critical field is given by

$$Hc = \Phi o/2\pi\xi^2 \quad \ldots\ldots\ldots\text{Eq. (A.1)}$$

From this equation the upper critical field with temperature dependence is given by,

$$Hc2(T) = Hc2(0)\left(\frac{1-t^2}{1+t^2}\right) \quad \ldots\ldots\ldots\text{Eq. (A.2)}$$

The upper critical field $H_{C2}$ versus temperature analysis yields an extrapolated zero-temperature value of approximately 11.7 T. This value, determined by fitting the $H_{C2}$-T phase diagram using conventional models such as the Ginzburg-Landau theory, indicates the magnetic field strength required to completely suppress superconductivity at 0 K. Nevertheless, conclusively attributing superconductivity to the compound remains challenging due to the significant electron-electron correlations induced by spin fluctuations, which can obscure the intrinsic superconducting behaviour. Complementary scanning tunnelling spectroscopy (STS) measurements were analysed using a d-wave pairing model based on Dynes' formula, as shown in Figure 13(d). The resulting fits reveal a superconducting energy gap of approximately 4.8 mev, significantly larger than the 3-4 mev gap commonly seen in conventional BCS superconductors. These results suggest that the pairing mechanism in $LiTi_2O_4$ is unconventional, potentially involving d-wave symmetry [39]. Overall, these observations emphasise the need for further detailed studies to unambiguously establish the superconducting nature of the compound.

Table 1: Summary of synthesis methods, characterization techniques, and superconducting properties for polycrystalline and epitaxial LTO samples.

| Reference | Sample Type | Synthesis Method | Measurement Techniques | Tc (K) | Superconducting Features | Additional Observations |
|---|---|---|---|---|---|---|
| [48] | Polycrystalline LTO | Solid-state reaction | Specific heat measurements | ~11.4 | Type-II, BCS-like, moderate coupling, fully gapped superconductor in the dirty limital metame | Estimated $Hc_2(0)$ ~ 11.7 T, Hc (0) ~ 0.32 T, $\xi_{GL}$ (0) ~ 55 Å, $\lambda_{GL}$ (0) ~ 1600 Å, $Hc_1(0)$ ~ 26 mT |
| [49] | Epitaxial LTO thin films | Pulsed laser deposition (PLD) | X-ray diffraction (XRD), resistivity, magnetization | ~13 | Superconductivity sensitive to structural defects; insights into microscopic mechanisms | First successful growth of epitaxial thin films; potential for device integration |
| [39] | Single-crystal LTO thin films | PLD | Magneto resistivity, upper critical field measurements | ~12.5 | Fourfold rotational symmetry in superconductivity, suggesting unconventional d-wave pairing | Intrinsic property of the superconducting phase; no anisotropy in the normal state |
| [50] | LTO (111) thin films | PLD | Scanning tunnelling microscopy (STM) | ~13 | Small superconducting energy gap, long coherence length at the surface | Presence of a pseudo gap at the Fermi energy modifying surface superconductivity |
| [51] | LTO (111) thin films | Epitaxial growth | Scanning tunnelling microscopy (STM) | ~13 | Triangular-shaped Abrikosov vortices, indicating electronic symmetry breaking under magnetic field | Josephson vortices observed along crystalline domain boundaries |
| [52] | LTO thin films | Epitaxial growth | Angle-resolved photoemission spectroscopy (ARPES) | ~13 | Abrupt flattening of near Fermi energy dispersion below 150 K, suggesting exotic quasiparticle states | Negative thermal expansion observed below 150 K |
| [53] | LTO thin films ([111], [110], [001] orientations) | PLD | Transport measurements, magnetoresistance | ~13 | Crystallographic orientation-dependent transport properties | Oxygen content significantly affects superconducting properties |
| [54] | Polycrystalline LTO | Solid-state reaction | Magnetoresistance measurements | ~13 | Nearly isotropic negative magnetoresistance at higher temperatures; prominently anisotropic positive magnetoresistance at lower temperatures | Indicates complex interplay between superconductivity and magnetoresistance |

From the findings of Fujisawa et al. the discovery of a novel quantum state of matter in the spinel oxide superconductor LTO, marked by a sudden flattening of electronic bands near the Fermi level below a characteristic temperature $T^*$ ~150 K, without any accompanying energy gap, band splitting, or structural distortion. A gapless Fermi surface persists across $T^*$, along with the emergence of negative thermal expansion (NTE), suggesting a distinct low-temperature phase. The state is attributed to a competition between orbital ordering tendencies and geometric frustration in the Ti-pyrochlore lattice. Future research directions include understanding the link between this anomalous metallic state and superconductivity, probing the microscopic nature of the transition at $T^*$, and exploiting thin-film and interface engineering to explore correlated physics in related systems. The findings also hold potential for designing thermally stable metallic electrodes for advanced nanoscale devices [52].

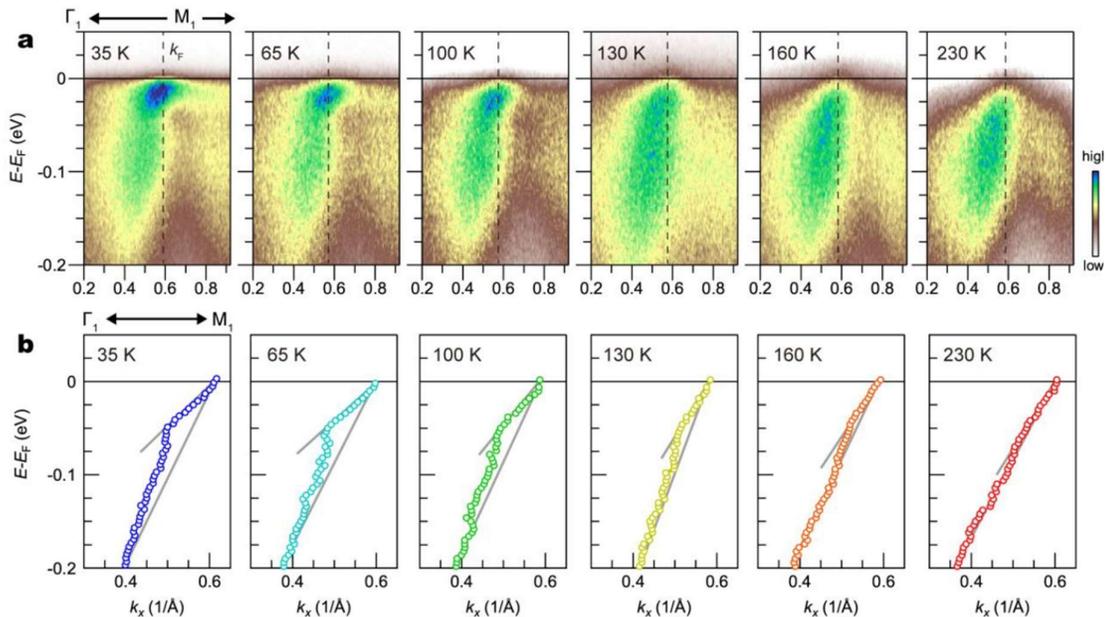

Figure 14. Temperature-dependent ARPES measurements of LTO along the $\Gamma_1$–$M_1$ direction. (a) ARPES intensity maps from 35 K to 230 K show a flat band near the Fermi level at low temperatures, which gradually becomes more dispersive with increasing temperature, indicating a loss of coherence. (b) Extracted band dispersions reveal a clear deviation from linearity below ~100 K, reflecting a renormalization effect likely tied to many-body

interactions. The Gray lines serve as linear references highlighting the temperature-induced band evolution [52].

From the graph (**Fig. 14**) ARPES intensity plots in panel (a) show the temperature-dependent band dispersion of LTO along the high-symmetry $\Gamma_1$–$M_1$ direction from 35 K to 230 K. The spectral intensity, represented by a colour scale (blue for high intensity and brown for low), reveals a distinct flat band-like feature near the Fermi level (E-$E_F$ = 0 eV, marked by a horizontal black line) at low temperatures (35-100 K), indicative of band renormalization or a kink structure. As the temperature increases, this feature gradually fades, and the band becomes more dispersive, with the spectral weight near the Fermi momentum $K_F$ (vertical dashed line) broadening, signifying a crossover from a coherent to an incoherent electronic state. Panel (b) presents the extracted band dispersions from the ARPES data, where the data points represent the peak positions in energy-momentum space. A clear flattening of the band is observed below ~100 K, suggesting many-body renormalization effects or enhanced electronic correlations. The Gray lines denote linear fits based on high-temperature dispersions, highlighting the nonlinearity and renormalization at lower temperatures. Together, these observations underscore a temperature-induced reconstruction of the low-energy electronic structure in LTO [52].

## 3.2 LVO: Heavy Fermion Behaviour

LVO is the first known d-electron heavy fermion system, making it a unique and extensively studied material. Heavy fermion behaviour is typically found in f-electron systems, where electrons interact strongly, leading to an extremely high effective mass [55, 56]. However, such behaviour in d-electron compounds is rare, making LVO an important example of how strong electron correlations can give rise to heavy fermion properties in transition metal oxides.

In heavy fermion systems, the effective electron mass increases due to interactions between localized and itinerant electrons. Two primary mechanisms are considered responsible for this in LVO. The first is Kondo-like screening, where localized magnetic moments are gradually screened by conduction electrons, forming a narrow quasiparticle resonance [57]. The second is Hund's coupling, which plays a crucial role in multi-orbital systems like LVO. Here, one set of orbitals ($a_{1g}$) becomes more localized, while another ($e_g^\pi$) remains itinerant. This orbital differentiation contributes to the formation of heavy quasiparticles, which means in LVO, Hund's coupling plays a crucial role in changing its electronic properties by inducing orbital variation, where the a1g orbitals become more localised while the e′g orbitals remain itinerant. This discrepancy leads to strong electron-electron interactions, particularly in the localised orbitals. This change enhances the effective mass of charge carriers. As a result, the system exhibits heavy quasiparticles, where electrons behave as though they have an increased effective mass due to strong interactions between localized and itinerant states. This heavy fermion-like behaviour is a signature of strongly correlated electron systems, with Hund's coupling playing a key role in enabling the coexistence of localised and itinerant electrons. These interactions significantly impact the material's transport and magnetic properties. Another important factor shaping the behaviour of LVO is geometric frustration. The vanadium atoms form a pyrochlore lattice, an arrangement that inhibits the formation of long-range magnetic order. Instead, spin fluctuations persist even at very low temperatures, further contributing to the enhanced effective mass of the charge carriers. This frustration helps stabilise the heavy fermion state and makes LVO different from conventional heavy fermion materials [58, 59].

The heavy fermion state is characterized by an enhanced electronic specific heat coefficient (γ) and a large effective mass $m^*$, both of which result from interactions between localized and itinerant electrons [60, 61]. The Sommerfeld coefficient (γ) is related to the density of states at the Fermi level $D(E_f)$ and the quasiparticle mass ($m$) by the equation [62]:

$$\gamma(T = 0) = \frac{\pi^2 K_B^2 D(E_f)}{3} \ldots\ldots\ldots \text{Eq. (B.1)}$$

where $K_B$ is the Boltzmann constant. The density of states is further linked to the effective mass as:

$$D(E_f) = \frac{m^* K_f V}{\pi^2 \hbar^2} \ldots\ldots\ldots \text{Eq. (B.2)}$$

where $K_f$ is the Fermi wavevector, given by:

$$K_f = \left(\frac{3\pi^2 N_e}{V}\right)^{1/3} \ldots\ldots\ldots \text{Eq. (B.3)}$$

where $N_e$ is the number of conduction electrons and V is the system volume [63]. These relations highlight how the effective mass enhancement in LVO is directly tied to its high electronic specific heat and electron density near the Fermi level.

The crystal structure of LVO also plays a significant role. It crystallizes in a normal spinel structure, where lithium occupies tetrahedral sites, and vanadium occupies octahedral sites. The $VO_6$ octahedra experience a slight trigonal distortion, which splits the $t_{2g}$ orbitals into a lower-energy $a_{1g}$ orbital and a higher-energy $e'g$ doublet. This splitting, combined with the mixed valence state of vanadium ($V^{3+}/V^{4+}$), leads to an unusual electronic structure that strongly influences the material's properties.

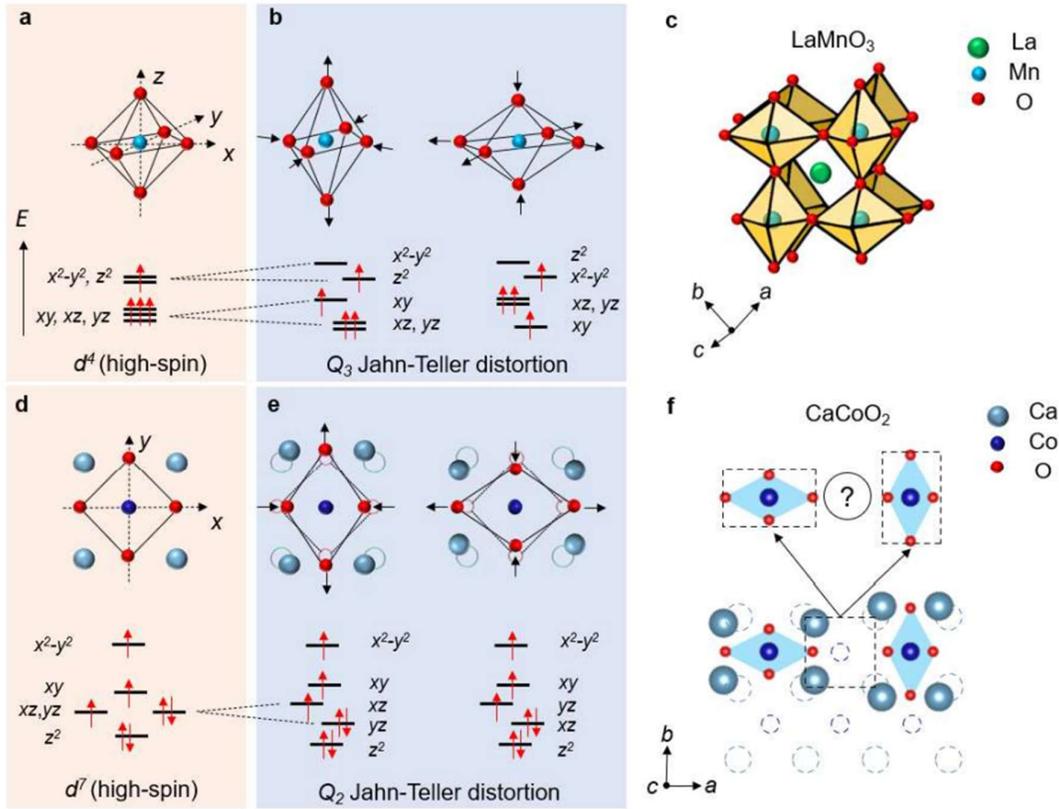

Figure 15. Jahn-Teller Distortions in 3D and 2D Oxide Lattices. (a) Transition metal octahedron (MO$_6$) with crystal field splitting for a high-spin d$^4$ configuration. (b) Octahedral distortions corresponding to a Q$_3$-type Jahn-Teller effect and resulting orbital energy changes. (c) Crystal structure of LaMnO$_3$ highlighting strong Q$_3$-type distortions from high-spin d$^4$ Mn$^{3+}$ ions, demonstrating a cooperative Jahn-Teller distortion. (d) Square-planar MO$_4$ unit with crystal field splitting for a high-spin d$^7$ configuration. (e) Square-planar distortions due to a Q$_2$-type Jahn-Teller effect and corresponding orbital level shifts. (f) Atomic displacements within the calcium layer causing geometric frustration, which inhibits the formation of cooperative Jahn-Teller distortions as seen in LaMnO$_3$ [64].

Following the explanation of crystal field splitting caused by trigonal distortion in the VO$_6$ octahedra and the presence of mixed valence vanadium ions, it is necessary to explore how these microscopic features influence observable physical properties. Low-temperature magnetic susceptibility, χ(0), provides insight into the system's magnetic response. In heavy fermion materials like LVO, the large effective mass of quasiparticles results in an enhanced Pauli-type susceptibility [65]. As a result, χ(0), the magnetic susceptibility at zero temperature, serves as a crucial indicator of the strength of electron correlations, reinforcing the interpretation of heavy fermion behaviour arising from the material's underlying electronic structure. One useful metric for quantifying these correlations is the Sommerfeld-Wilson ratio (RW), which expresses the relationship between the electronic specific heat coefficient (γ) and the magnetic susceptibility χ (0). It is defined as:[66],

$$R_w = \frac{2\pi^2 K_B^2 \chi(0)}{3\mu_{eff}^2 \gamma(0)} \dots\dots\dots \text{Eq. (B.4)}$$

where $\chi(0)$ is the low-temperature magnetic susceptibility, and $\mu_{eff}$ is the effective magnetic moment. This ratio further characterizes the heavy fermion behaviour by linking magnetic properties to the quasiparticle mass enhancement.

The $T^2$ dependence explicitly appears in the expression for resistivity in heavy fermion systems and is central to the Kadowaki-Woods ratio. In these systems, the low-temperature resistivity behaves as $\rho(T) = \rho_0 + AT^2$, where A is the coefficient of the $T^2$ term. The Kadowaki-Woods ratio, $\frac{A}{\gamma^2}$, relates this transport property to the square of the electronic specific heat coefficient, $\gamma$ [67].

$$\frac{A}{\gamma^2} \approx 1.0 \times 10^{-5} \mu\Omega cm (molK^2)^2/(mJ)^2 \ldots\ldots\ldots \text{Eq. (B.5)}$$

This empirical relation is used to characterize the strength of electron-electron correlations, with the ratio remaining roughly constant across many heavy fermion materials. Therefore, while the Sommerfeld-Wilson ratio links magnetism and thermodynamics, the Kadowaki-Woods ratio connects thermal and electrical transport, and it is here where the $T^2$ behaviour of resistivity plays a critical role. The fact that LVO follows this empirical relation further confirms its heavy fermion nature.

Electronic structure calculations, including Density Functional Theory (DFT) and Dynamical Mean Field Theory (DMFT), reveal that the V-3d states in LVO split into two distinct sets of bands. The $a_1g$ band, characterised by a narrow bandwidth, is strongly correlated and tends toward localisation, whereas the broader $e'g$ band supports itinerant electron motion. This separation is crucial for understanding the origin of heavy fermion behaviour in LVO. The renormalisation of the narrow $a_1g$ band results in a sharp quasiparticle peak near the Fermi level, a feature that has been confirmed through photoemission spectroscopy, highlighting the presence of strong electronic correlations.

The emergence of the heavy fermion state in LVO is driven by a combination of closely linked factors. Strong electron correlations primarily affect the narrow $a_1g$ band, significantly enhancing the effective mass of the quasiparticles. Hund's coupling favours high-spin configurations, stabilising localised moments in the $a_1g$ orbital while allowing $e'g$ electrons to remain itinerant. This dynamic gives rise to an orbital-selective behaviour, placing LVO near a Mott transition, specifically an orbital-selective Mott state where localised and itinerant electrons coexist. Geometric frustration, inherent to the pyrochlore lattice formed by vanadium atoms, suppresses long-range magnetic ordering and promotes persistent spin fluctuations. These fluctuations further renormalize the quasiparticles and deepen the heavy fermion character. Experimental evidence strongly supports this theoretical framework. LVO exhibits a remarkably large electronic specific heat coefficient ($\gamma$), a defining feature of heavy fermion systems. Its magnetic susceptibility follows Curie-Weiss behaviour at high temperatures, indicating the presence of local moments, but becomes nearly temperature-independent at low temperatures, characteristic of Kondo-like screening. Additionally, photoemission spectroscopy reveals a sharp quasiparticle peak near the Fermi level, underscoring the presence of strong electron correlations. Together, these findings establish LVO as an exceptional example of a d-electron heavy fermion material, where the interplay of orbital physics, strong correlations, and geometric frustration dictates its unusual electronic behaviour.

The large electronic specific heat coefficient ($\gamma \approx 0.4$ J/mol·K$^2$) indicates the presence of heavy quasiparticles, suggesting that the effective mass of the electrons exceeds 60 times that of a free electron, consistent with the hallmarks of a heavy fermion system [68]. This high value points to pronounced mass renormalisation, a defining feature of heavy-fermion systems. Additionally, research suggests that geometric frustration in LVO gives rise to strong spin and orbital fluctuations, which are key to understanding its heavy-fermion behaviour. These fluctuations, in combination with the mixed valence state of vanadium ions, create a complex interplay between localized and itinerant electrons, resulting in the material's distinctive electronic characteristics.

The nearly temperature-independent magnetic susceptibility observed at low temperatures indicates the screening of local moments, consistent with a Kondo-like ground state. Furthermore, the magnetic frustration inherent to the pyrochlore lattice of LVO prevents long-range magnetic ordering, promoting a landscape of nearly degenerate states that underpins its unconventional magnetic behaviour [69].

A sharp quasiparticle peak appearing just above the Fermi level at low temperatures reflects the onset of orbital-selective correlations, further supporting the presence of Kondo resonance behaviour in the system [68]. This observation is pivotal in unravelling the electronic structure of LVO, particularly the delicate interplay between localised and itinerant electron states. While these findings strongly support the heavy-fermion model, alternative interpretations have also emerged. Some studies propose that those competing interactions, especially geometric frustration inherent in the pyrochlore lattice, may suppress conventional magnetic ordering and instead give rise to a metallic spin-liquid state. This highlights the complexity of LVO's ground state and the need for further investigation to fully understand the nature of its correlated electron behaviour [70].

Table 2: Summary of key experimental studies on polycrystalline LVO highlighting measurement techniques, heavy fermion characteristics, and additional physical insights.

| Reference | Sample Type | Measurement Techniques | Heavy Fermion Features | Additional Observations |
|---|---|---|---|---|
| [55] | Polycrystalline LVO | Specific heat, magnetic susceptibility, resistivity | Very large Sommerfeld coefficient ($\gamma \approx 420$ mJ/mol·K$^2$); clear Fermi liquid behaviour below T~20K | No magnetic ordering down to very low T~20K; first report of a d-electron heavy fermion system |
| [2] | Polycrystalline LVO | Specific heat, resistivity, susceptibility | Enhanced effective mass and heavy quasiparticles; Kondo-like temperature dependence | Crossover from high-T Curie–Weiss behaviour to low-T Fermi liquid state |
| [71] | Polycrystalline LVO | Neutron scattering, magnetic susceptibility | Pronounced spin fluctuations; signatures of heavy fermion renormalization | Geometric frustration in the spinel structure is suggested to influence spin dynamics |
| [72] | Polycrystalline LVO | Inelastic neutron scattering | Antiferromagnetic spin fluctuations correlating with the heavy fermion state | Dynamic spin correlations persist in the absence of static order |
| [73] | Polycrystalline LVO | Muon spin relaxation (μSR) | Spin fluctuation rate shows linear temperature dependence; absence of static magnetism persists | μSR reanalysis reveals quasi-1D spin dynamics consistent with inelastic neutron scattering. Suggests heavy quasiparticle formation via geometrically constrained t$_{2g}$ band and 1D-to-3D crossover. |
| [65] | Polycrystalline LVO | Resistivity (under pressure), | Intrinsic susceptibility χ(T) shows T-independent behaviour below ~30 K with a broad | Heavy fermion state arises from antiferromagnetic V–V exchange (J/k$_B$ ≈ 20 K); not explained by |

| | | magnetic susceptibility | maximum at ~16 K; Curie-Weiss behaviour above 50–100 K; no spin-glass ordering observed | crystal field, Kondo, or Coqblin-Schrieffer models; sensitive to lattice tuning and impurities |

### 3.3 MTO: Orbital Ordering and Electron-Lattice Coupling

MTO is one of the transitions-metal spinels that has attracted considerable attention because of its sharp metal-insulator transition (MIT) accompanied by a structural distortion, orbital ordering, and spin dimerization. In the high-temperature cubic phase, the Ti ions are in the 3+ oxidation state with a $d^1$ electronic configuration. A single d-electron occupies one of the three $t_{2g}$ orbitals ($d_{xy}$, $d_{xz}$, $d_{yz}$), and in the ideal cubic environment, these orbitals are triply degenerate. However, this degeneracy is inherently unstable according to the Jahn–Teller theorem [74, 75], which states that any nonlinear molecule (or crystal complex) in a degenerate electronic state will undergo distortion to lower its symmetry and energy [76].

The *Jahn–Teller effect* in transition–metal compounds, such as MTO, is a cooperative phenomenon. In an isolated $TiO_6$ octahedron, the degeneracy of the $t_{2g}$ levels can be partially lifted by a local distortion; however, in a solid, the distortions on the individual octahedra interact. When many $TiO_6$ octahedra are distorted in a coherent fashion, the crystal symmetry is reduced from cubic to tetragonal. This symmetry lowering has two main consequences.

- *Orbital Selection:* The distortion splits the $t_{2g}$ manifold such that one orbital (or a pair of orbitals) is lowered in energy relative to the others. For MTO, theoretical studies, density functional theory, and Hubbard U (LDA+U) calculations indicate that the $d_{xz}$ and $d_{yz}$ orbitals tend to be stabilized relative to the $d_{xy}$ orbital. This orbital ordering effectively polarizes the electronic state and favours electron occupation in certain crystallographic directions [75, 77].
- *Dimensional Reduction:* With preferential occupation of orbitals that have lobes oriented along specific bonds, the original three-dimensional electron hopping network is reorganized into quasi one-dimensional pathways. For example, if the $d_{xz}$ and $d_{yz}$ orbitals become predominantly occupied, dominant electron hopping may occur along chains running in directions in which the overlap between these orbitals is maximized [77].

Mathematically, in a simple tight-binding picture the dispersion relation along a given chain can be written as

$$E(k) = -2t \cos(k) \ldots\ldots\ldots \text{Eq. (C.1)}$$

where $t$ is the effective hopping integral, and $k$ is the wave vector along the chain direction. The one-dimensional nature of the resulting bands is crucial because 1D systems are particularly vulnerable to instabilities, such as Peierls distortion.

*Peierls Instability and Formation of Spin Dimers*

In a one-dimensional metal, the Fermi surface consists of two points, and electron-lattice coupling can drive a spontaneous lattice distortion if there is a perfect nesting condition. In case of MTO, once orbital ordering reduces the effective dimensionality of the conduction bands, the partially filled 1D band becomes unstable against Peierls distortion. For a chain with a given electron filling (e.g., quarter filling when only one orbital per Ti ion is active), the Fermi wave vector is approximately $k_F \approx \pi/2$. In such a scenario, the nesting condition is satisfied for modulation with wave vector $Q = 2k_F \approx \pi$. This leads to periodic modulation of the lattice with doubling of the unit cell along the chain.

Peierls distortion results in alternating short and long bonds along the chain. On short bonds, the overlap between the orbitals increases significantly, which enhances the formation of a bonding molecular orbital; the electrons on these bonds can form spin singlets (dimers). The energy gain associated with the formation of a spin singlet on a dimer is of the order of the exchange energy and can be estimated using super exchange theory. Furthermore, the opening of an energy gap $\Delta$ at the Fermi level due to the dimerization can be approximated by $\Delta \sim 2t\,\delta$, where $\delta$ is a measure of the bond-length modulation amplitude (i.e. the distortion parameter). This gap is responsible for the insulating behaviour at low temperature.

Thus, the Peierls mechanism driven by the quasi-1D band structure induced by orbital ordering leads to both a lattice dimerization and the formation of spin singlet dimers. The magnetic susceptibility drops sharply below the transition temperature because the paired spins form nonmagnetic singlet states, while the gap in the electronic spectrum drives the system into an insulating state. Below approximately 260 K, MTO experiences an orbital ordering transition, which is crucial for understanding its metal-insulator transition and associated phenomena. Below 260 K, the titanium $t_{2g}$ orbitals in MTO arrange into an ordered configuration, minimizing the system's energy and stabilizing its electronic structure. This ordering lifts orbital degeneracy and enhances stability [78]. The orbital ordering induces Peierls-like distortion, resulting in the dimerization of Ti-Ti pairs, which reduces lattice symmetry and opens a charge gap [79]. This structural distortion facilitates a transition from a metallic state at high temperatures to an insulating state at low temperatures, with significant changes in electronic properties observed around 250 K [80].

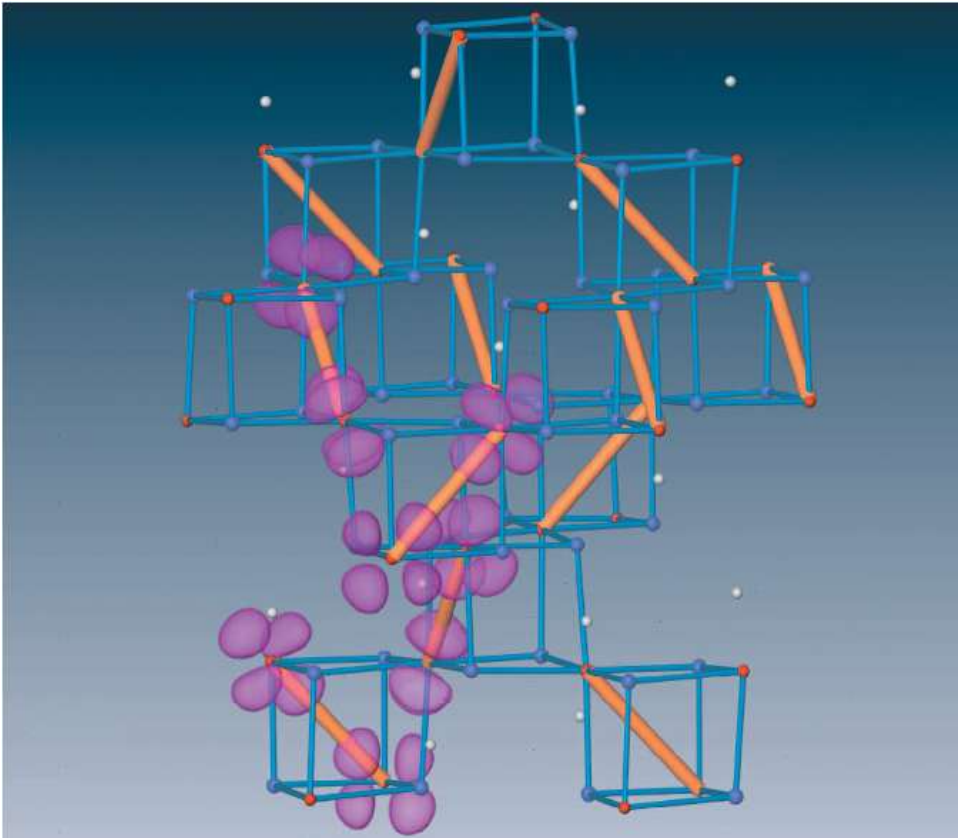

Figure 16. Orbital ordering in the tetragonal crystal structure of MTO calculated using the LDA+U method with U = 3 eV. The electron density of the most occupied $t_{2g}$ orbital is shown. Lattice parameters are adopted from [32]. Orange lines indicate the shortest Ti-Ti bonds in the distorted lattice, with Ti atoms shown in red, O in blue, and Mg in white [81].

The phenomena in MTO are interdependent. In this picture, the metal insulator transition is not a simple order-disorder transition but rather a cooperative phenomenon where the local electronic correlations (which drive the Jahn–Teller distortion and orbital ordering) and the lattice instability (the Peierls distortion) act together [82]. The structural dimerization observed experimentally often seen as a helical pattern of alternating short and long Ti-Ti bonds is a direct consequence of this interplay [83]. Spin dimerization is inherently linked to the orbital ordering because the spatial orientation of the occupied orbitals determines which bonds are strengthened; the short bonds are those along which the orbital overlap is maximized, allowing for the formation of robust spin singlets. Advanced theoretical models, including LDA+U calculations, have confirmed that the insulating state in MTO is characterized by a strong orbital polarization and a significant splitting between bonding and antibonding states along the dimerized chains. These calculations reveal that for an appropriate choice of the Hubbard U parameter, the system undergoes an orbital selective metal insulator transition. In this transition, one set of orbitals (the ones involved in the quasi 1D chains) drives the gap opening, while the other orbitals are pushed to higher energies, further reinforcing the orbital order [84].

The above-mentioned behaviours are supported by experimental results from the following review of literature.

## Table 3: Overview of Orbital Ordering Characteristics in Polycrystalline MTO

| Reference | Sample | Method of preparation | Measurement Techniques | Temperature (K) | Orbital Ordering Features |
|---|---|---|---|---|---|
| [85] | Polycrystalline MTO | Solid state reaction | Crystallographic investigations | ~260 | One-dimensional orbital wave ordering leading to dimer formation |
| [32] | Polycrystalline MTO | Solid state reaction | Band structure calculations | -- | Orbitally ordered band insulator with dimerized Ti-Ti distances |
| [86] | Polycrystalline MTO | Solid state reaction | Thermal conductivity measurements | -- | Rapidly fluctuating orbital occupancy leading to an orbital-liquid state |
| [87] | Polycrystalline MTO | Solid state reaction | Spin-orbital Hamiltonian analysis | -- | Formation of a valence bond crystal with orbital ordering; dimers along helical chains |
| [88] | Polycrystalline MTO | Solid state reaction | X-ray and neutron pair distribution function analyses | Up to 500 K | Local tetragonality in the metallic phase; fluctuating two-orbital $t_{2g}$ orbital-degeneracy-lifted precursor state |
| [33] | Polycrystalline MTO | Solid state reaction | Optical measurements | 240 | 1D Peierls transition driven by the ordering of $d_{YZ}$ and $d_{ZX}$ orbitals |
| [89] | Polycrystalline MTO | Solid state reaction | Theoretical analysis | -- | Orbital ordering phenomena in d- and f- electron systems |
| [76] | Polycrystalline MTO | Solid state reaction | Theoretical analysis | -- | Orbital physics: glorious past, bright future |
| [90] | Thin film MTO | Epitaxial thin films grown by PLD | Spectroscopic measurements | -- | Competition between orbital order and superconductivity |
| [91] | Thin film MTO | --- | Not specified | ~3 | Superconducting transition with suppressed orbital order gap |

The strong electron-lattice coupling in MTO is evidenced by the gradual formation of a magnetic spin-singlet state and orbital density wave as temperature decreases [80]. Local distortions persist even at high temperatures, indicating a complex interplay between local and global structural changes [79]. Conversely, in MTO, electron-lattice coupling and orbital ordering play key roles in driving its insulating behaviour. However, when electron doping is increased, superconductivity emerges, indicating a competing mechanism that can disrupt the established ordered state. The dimerization of titanium ions leads to the formation of spin-singlet pairs, which results in a non-magnetic ground state [92]. This behaviour is seen across several titanium oxides, highlighting how strong electron correlations help stabilize the spin-singlet state [93]. Strong electron-lattice coupling plays a crucial role in the formation of charge order and spin-singlet pairs, profoundly influencing the material's properties [92, 93]. The interaction between orbital and spin degrees of freedom is pivotal, as it shapes both the structural transitions and the overall energy landscape of the material [94].

At low temperatures, the opening of a charge gap results in a noticeable drop in electrical conductivity, signaling a transition to the insulating state [93]. This transition offers valuable insights into the mechanisms governing strongly correlated electron systems, emphasizing the crucial role of electron-lattice interactions in shaping their physical properties [95].

## 3. Applications and Technological Implications

Spinel oxides exhibit a broad spectrum of applications and technological relevance across materials science, energy storage, and catalysis. Their unique structural flexibility and stability underpin advancements in high-temperature lubrication, electrocatalysis, and infrared radiation management, highlighting their versatility and potential for future innovation.

High-Temperature Lubrication: Spinel oxides can form thermally stable, self-lubricating layers on superalloys, maintaining low coefficients of friction (0.10–0.32) at elevated temperatures ranging from 600°C to 900°C. This makes them ideal for aerospace, turbine, and high-performance mechanical applications [96]. This property is crucial for enhancing the wear resistance of metals in oxidative environments, where conventional lubricants degrade or fail, making spinels ideal for aerospace, turbine, and high-performance mechanical applications.

Electrocatalytic Applications: Spinel-type metal oxides, such as $NiCo_2O_4$, exhibit remarkable electrocatalytic performance in water splitting reactions, particularly in the oxygen evolution reaction (OER). Their mixed-valence states and high electrical conductivity contribute to low overpotentials and high current densities, making them attractive alternatives to noble metal-based catalysts. Additionally, their structural robustness and cost-effectiveness enhance their suitability for sustainable energy technologies such as electrolysers and metal-air batteries. [97]. Creating amorphous/crystalline heterointerfaces in spinel oxides significantly boosts their electrocatalytic performance, making them promising and affordable alternatives to traditional noble metal catalysts.

Energy Storage and Conductivity: Spinel oxides hold great promise for energy storage applications, and recent advances have even used machine learning to predict their electronic conductivity based on their composition, helping to design more efficient materials [98]. High nickel content in spinel oxides is linked to better electrical conductivity, which is crucial for enhancing the performance of batteries and supercapacitors.

Infrared Radiation Management: High-entropy spinel oxides stand out for their strong infrared emissivity and excellent thermal stability, making them ideal candidates for energy-efficient solutions in industrial furnace applications [99].

Although spinel oxides have many advantages, it's still challenging to optimize their properties for specific uses, especially when trying to balance high activity with long-term stability in electrocatalysis [100]. Continued research is essential to unlock their full potential across various technological domains.

The spinel oxides LTO, LVO, and MTO exhibit unique electronic properties, making them promising candidates for various advanced applications. Below is an overview of their potential applications, supported by recent research findings:

### 4.1 LTO:

Superconducting Electronics: LTO is notable as the only known oxide spinel superconductor, with a critical temperature ($T_c$) of approximately 13.7 K. Its superconducting properties are sensitive to lattice strain and microstructural disorder, which can be systematically studied using epitaxial thin films. These traits make LTO a strong contender for use in superconducting electronic devices [101].

Spintronics and Quantum Devices: Research has shown unusual magnetoresistance in LTO, linked to spin fluctuations and orbital effects. This points to its potential for spintronic technologies and quantum devices, where managing spin-dependent transport is key [102].

Energy Storage: Thanks to its high lithium diffusivity and strong structural stability during lithiation, LTO stands out as a promising electrode material for lithium-ion batteries. Its minimal changes in lattice parameters during charging and discharging help ensure a long and reliable cycle life.

### 4.2 LVO:

Heavy Fermion Systems: LVO is unique among oxides for exhibiting heavy fermion behaviour, characterized by a large electronic specific heat coefficient. This property makes it an interesting system for studying strong electron correlations and potential applications in low-temperature physics [103].

Battery Applications: $Li_3VO_4$, a related compound, has been explored as an anode material for lithium-ion batteries. Carbon composites of $Li_3VO_4$ synthesized via Sol-Gel methods have demonstrated enhanced electrochemical performance, indicating potential for energy storage applications [104].

### 4.3 MTO:

Superconductivity: Recent studies have reported superconductivity in V-doped, Mg-deficient MTO, with a $T_c$ of approximately 16 K. This superconducting phase is stabilized as a thin surface layer on top of the insulating bulk, suggesting potential for surface-based superconducting applications [91].

Mechanism Insights: Superconductivity in V-doped MTO appears to arise from multiple charge transfers and the reduction of Jahn-Teller distortions, which together weaken the material's Mott insulating state. Gaining a deeper understanding of these processes can help pave the way for designing new superconductors [105].

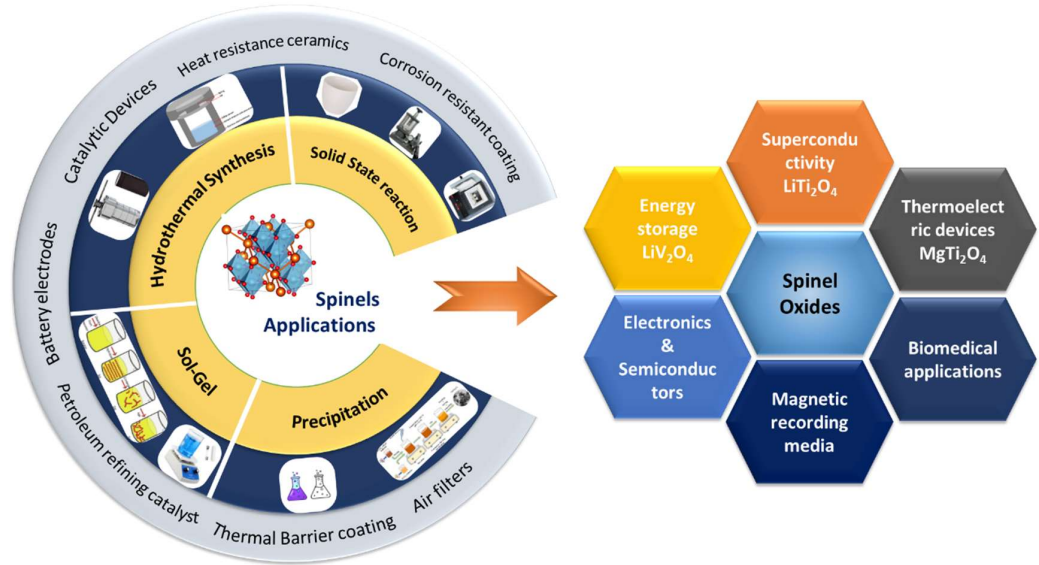

Figure 17. Schematic overview of the primary synthetic routes for spinel materials hydrothermal, sol-gel, precipitation, and solid-state reactions and their broad range of industrial and technological applications. The left section highlights examples such as catalytic devices, battery electrodes, and thermal barrier coatings, while the arrow points to notable spinel oxides used in superconductivity, thermoelectric devices, energy storage, electronics, and biomedical applications[106-108].

4. **Future Direction**

Despite notable advances, several challenges still limit the full exploitation of LTO, LVO, and MTO, both in understanding their fundamental physics and in developing practical devices. Overcoming these hurdles is essential to unlock their unique electronic behaviours for future quantum and spintronic technologies. LTO, a rare oxide superconductor, faces reproducibility issues in synthesis due to lithium out-diffusion, oxygen non-stoichiometry, and phase instability. These factors negatively impact its superconducting properties and critical temperature. Additionally, the unconventional nature of its superconductivity and the origin of pseudo gap-like features observed at higher temperatures remain unclear. It is crucial to determine whether these anomalies arise from preformed Cooper pairs, quantum metallicity, or other exotic correlations. Moving forward, research should aim to probe the onset and spatial fluctuations of superconductivity in LTO, with advanced techniques like low-frequency noise spectroscopy that can sensitively detect inhomogeneities and phase fluctuations [109-113].

LVO, known for its unique heavy fermionic behaviour among oxides, presents challenges in disentangling the effects of strong electron correlations, geometrical frustration, and orbital-selective Mott physics. Furthermore, the sample quality, especially in polycrystalline forms, limits the ability to extract intrinsic behaviours. Understanding whether the heavy-fermion state coexists with local orbital ordering or fluctuating magnetic states requires systematic ARPES, NMR, and noise measurements, which can sensitively probe the low-energy excitations and dynamic instabilities.

MTO exhibits a sharp MIT and orbital ordering, which are highly sensitive to subtle Structural distortions, strain effects, and variations in stoichiometry play crucial roles in these materials. However, the precise relationship between orbital ordering and the metal-insulator transition (MIT), along with how this relationship can be tuned through epitaxial strain or chemical doping, is still not fully understood. Employing temporal noise analysis, particularly across the transition region, could reveal important details such as critical slowing down, domain dynamics, and phase coexistence. This approach offers a unique and powerful way to investigate the complex ordering phenomena in these systems [114].

To address these open issues, future research should prioritize:

- Growth of high-quality thin films and single crystals using advanced techniques such as molecular beam epitaxy (MBE), PLD, and oxide molecular beam epitaxy, with precise control over stoichiometry and strain.
- In situ characterizations using ARPES, STM, neutron/X-ray scattering, and local noise measurements to probe fluctuations, ordering, and inhomogeneities at mesoscopic and atomic scales.
- Theoretical modelling using many-body techniques such as DFT+DMFT, cluster extensions, and orbital-selective modelling frameworks to capture the entangled spin-orbital-lattice interactions and simulate their response to external fields or confinement.

## 5. Conclusion

To sum up, spinel oxides offer an exciting playground to explore how structure, electron interactions, and new physical phenomena all come together. In materials like LVO, LTO, and MTO, the delicate balance between electron correlations, lattice effects, and orbital behaviour pushes us beyond traditional theories and reveals fascinating properties, ranging from heavy fermion behaviour and unconventional superconductivity to metal-insulator transitions driven by orbital ordering. The unique geometry of the spinel structure plays a key role here, giving us ways to tune material properties by carefully controlling how cations are arranged or by applying external influences. Beyond the science, these materials hold great promise for practical uses, whether in energy storage, electrocatalysis, or even high-temperature lubrication. By combining deep theoretical understanding with experimental advances, this field not only expands our knowledge of complex electron systems but also lays the groundwork for creating next-generation materials to meet the challenges of future technologies.

## 6. Acknowledgement

The authors gratefully acknowledge the National Institute of Technology Rourkela for providing the necessary laboratory infrastructure and research facilities that supported the preparation of this review.

## 7. References

1. Walsh, A., et al., *Structural, magnetic, and electronic properties of the Co-Fe-Al oxide spinel system: Density-functional theory calculations.* Physical Review B—Condensed Matter and Materials Physics, 2007. **76**(16): p. 165119. https://doi.org/10.1103/PhysRevB.76.165119

2. Urano, C., et al., *LiV 2 O 4 spinel as a heavy-mass Fermi liquid: Anomalous transport and role of geometrical frustration.* Physical review letters, 2000. **85**(5): p. 1052. https://doi.org/10.1103/PhysRevB.76.165119
3. Harrison, M., P. Edwards, and J. Goodenough, *The superconductor-semiconductor transition in the Li1+ xTi2-xO4 spinel system.* Philosophical Magazine B, 1985. **52**(3): p. 679-699. https://doi.org/10.1103/PhysRevB.76.165119
4. Garlea, V.O., et al., *Magnetic and orbital ordering in the spinel MnV 2 O 4.* Physical review letters, 2008. **100**(6): p. 066404. https://doi.org/10.1103/PhysRevLett.100.066404
5. Tristan, N., et al., *Geometric frustration in the cubic spinels M Al 2 O 4 (M= Co, Fe, and Mn).* Physical Review B—Condensed Matter and Materials Physics, 2005. **72**(17): p. 174404. https://doi.org/10.1103/PhysRevB.72.174404
6. Büttgen, N., et al., *Quantum criticality in transition-metal oxides.* Journal of Low Temperature Physics, 2010. **161**: p. 148-166. https://doi.org/10.1007/s10909-010-0200-9
7. Shao, Z., et al., *Synthesis and advantages of spinel-type composites.* Materials Chemistry Frontiers, 2023. **7**(21): p. 5288-5308. https://doi.org/10.1039/D3QM00416C
8. Al-Rubaye, S.H.K., *Nanostructured electrodes based on spinel oxide composites for next generation energy storage devices*. 2018, University of Wollongong. https://ro.uow.edu.au/theses1
9. Hou, S., et al., *Unlocking the origins of highly reversible lithium storage and stable cycling in a spinel high-entropy oxide anode for lithium-ion batteries.* Advanced Functional Materials, 2024. **34**(4): p. 2307923. https://doi.org/10.1002/adfm.202307923
10. Biagioni, C. and M. Pasero, *The systematics of the spinel-type minerals: An overview.* American Mineralogist, 2014. **99**(7): p. 1254-1264. http://dx.doi.org/10.2138/am.2014.4816
11. Sickafus, K.E., J.M. Wills, and N.W. Grimes, *Structure of spinel.* Journal of the American Ceramic Society, 1999. **82**(12): p. 3279-3292. https://doi.org/10.1111/j.1151-2916.1999.tb02241.x
12. Das, D., R. Biswas, and S. Ghosh, *Systematic analysis of structural and magnetic properties of spinel CoB2O4 (B= Cr, Mn and Fe) compounds from their electronic structures.* Journal of Physics: Condensed Matter, 2016. **28**(44): p. 446001. https://doi.org/10.1088/0953-8984/28/44/446001
13. Chaibeddra, D., et al., *Synthesis, physical and electrochemical characterization of CoCr2O4 and its application as photocatalyst under solar irradiation.* Inorganic Chemistry Communications, 2023. **155**: p. 111116. https://doi.org/10.1016/j.inoche.2023.111116
14. Ma, Y., et al., *Sandwich-shell structured CoMn2O4/C hollow nanospheres for performance-enhanced sodium-ion hybrid supercapacitor.* Advanced Energy Materials, 2022. **12**(11): p. 2103820. https://doi.org/10.1002/aenm.202103820
15. Sharifianjazi, F., et al., *Magnetic CoFe2O4 nanoparticles doped with metal ions: a review.* Ceramics International, 2020. **46**(11): p. 18391-18412. https://doi.org/10.1016/j.ceramint.2020.04.202
16. Mo, S.-D. and W. Ching, *Electronic structure of normal, inverse, and partially inverse spinels in the MgAl 2 O 4 system.* Physical Review B, 1996. **54**(23): p. 16555. https://doi.org/10.1103/PhysRevB.54.16555
17. Shannon, N., *Mixed valence on a pyrochlore lattice - LiV2O4 as a geometrically frustrated magnet*, in *Strongly Correlated Electrons*. 2002, Cornell University. https://doi.org/10.1140/epjb/e2002-00186-9
18. Yamashita, Y. and K. Ueda, *Spin-orbital fluctuations and a large mass enhancement in ${\mathrm{LiV}}_{2}{\mathrm{O}}_{4}$.* Physical Review B, 2003. **67**(19): p. 195107. https://doi.org/10.1103/PhysRevB.67.195107
19. Tsunetsugu, H., *Instability in the t 2g-Band Model on the Pyrochlore Lattice.* Journal of the Physical Society of Japan, 2002. **71**(8): p. 1844-1847. https://doi.org/10.1143/JPSJ.71.1844

20. Takagi, H., et al., *Transport properties of metallic LiV2O4 single crystals—heavy mass Fermi liquid behavior.* Materials Science and Engineering: B, 1999. **63**(1-2): p. 147-150. https://doi.org/10.1016/S0921-5107(99)00065-3
21. LI, G., et al., *Synthesis and characterization of LiV2O4 as the cathode in secondary lithium batteries.* Denki Kagaku oyobi Kogyo Butsuri Kagaku, 1996. **64**(3): p. 202-206. https://doi.org/10.5796/kogyobutsurikagaku.64.202
22. Shimizu, Y., et al., *An orbital-selective spin liquid in a frustrated heavy fermion spinel LiV2O4.* Nature Communications, 2012. **3**(1): p. 981. https://doi.org/10.1038/ncomms1979
23. Onoda, M., *Mixed-valence ion model for incoherent metallic state in heavy-fermion spinel LiV2O4.* Journal of the Physical Society of Japan, 2021. **90**(3): p. 034701. https://doi.org/10.7566/JPSJ.90.034701
24. Nekrasov, I.A., et al., *Orbital state and magnetic properties of LiV 2 O 4.* Physical Review B, 2003. **67**(8): p. 085111. https://doi.org/10.1103/PhysRevB.67.085111
25. Okabe, H., et al., *Metallic spin-liquid-like behavior of ${\mathrm{LiV}}_{2}{\mathrm{O}}_{4}$.* Physical Review B, 2019. **99**(4): p. 041113. https://doi.org/10.1103/PhysRevB.99.041113
26. Satpathy, S. and R.M. Martin, *Electronic structure of the superconducting oxide spinel Li${\mathrm{Ti}}_{2}$${\mathrm{O}}_{4}$.* Physical Review B, 1987. **36**(13): p. 7269-7272. https://doi.org/10.1103/PhysRevB.36.7269
27. Johnston, D.C., *Superconducting and normal state properties of Li1+xTi2−xO4 spinel compounds. I. Preparation, crystallography, superconducting properties, electrical resistivity, dielectric behavior, and magnetic susceptibility.* Journal of Low Temperature Physics, 1976. **25**(1): p. 145-175. https://doi.org/10.1007/BF00654827
28. Wei, Z., et al., *Two superconductor-insulator phase transitions in the spinel oxide Li 1±x Ti 2 O 4− δ induced by ionic liquid gating.* Physical Review B, 2021. **103**(14): p. L140501. https://doi.org/10.1103/PhysRevB.103.L140501
29. Edwards, P.P., et al., *A study of the spinel materials LiTi2O4 and Li43Ti53O4 by photoelectron spectroscopy.* Journal of Solid State Chemistry, 1984. **54**(2): p. 127-135. https://doi.org/10.1016/0022-4596(84)90140-3
30. Ohsawa, T., et al., *Origin of Optical Transparency in a Transparent Superconductor LiTi2O4.* ACS Applied Electronic Materials, 2020. **2**(2): p. 517-522. https://pubs.acs.org/doi/10.1021/acsaelm.9b00751
31. Takahashi, Y., Y. Gotoh, and J. Akimoto, *Structure and electron density analysis of oxide spinel LiTi2O4.* Journal of Physics and Chemistry of Solids, 2002. **63**(6-8): p. 987-990. https://doi.org/10.1016/S0022-3697(02)00105-1
32. Schmidt, M., et al., *Spin Singlet Formation in ${\mathrm{M}}\mathrm{g}\mathrm{T}\mathrm{i}}_{2}{\mathrm{O}}_{4}$: Evidence of a Helical Dimerization Pattern.* Physical Review Letters, 2004. **92**(5): p. 056402. https://doi.org/10.1103/PhysRevLett.92.056402
33. Zhou, J., et al., *Optical study of ${\mathrm{MgTi}}_{2}{\mathrm{O}}_{4}$: Evidence for an orbital-Peierls state.* Physical Review B, 2006. **74**(24): p. 245102. https://doi.org/10.1103/PhysRevB.74.245102
34. Khomskii, D.I. and T. Mizokawa, *Orbitally Induced Peierls State in Spinels.* Physical Review Letters, 2005. **94**(15): p. 156402. https://doi.org/10.1103/PhysRevLett.94.156402
35. Oda, T., et al., *Electron-phonon interaction, lattice dynamics and superconductivity of an oxide spinel LiTi2O4.* Journal of Physics: Condensed Matter, 1994. **6**(35): p. 6997. https://doi.org/10.1088/0953-8984/6/35/009
36. Imbihl, R., *density of states*, in *Catalysis from A to Z*. https://doi.org/10.1002/9783527809080.
37. Wartak, M.S. and C.-Y. Fong, *Field Guide to Solid State Physics - Bragg's Law*. 2019, SPIE. https://ui.adsabs.harvard.edu/link_gateway/2019fgss.book.....W/doi:10.1117/3.2510243

38. Canadell, E., et al., *181Density of states*, in *Orbital Approach to the Electronic Structure of Solids*. 2012, Oxford University Press. p. 0. https://doi.org/10.1093/acprof:oso/9780199534937.003.0010
39. Xue, H., et al., *Fourfold Symmetric Superconductivity in Spinel Oxide LiTi2O4(001) Thin Films.* ACS Nano, 2022. **16**(11): p. 19464-19471. https://doi.org/10.1021/acsnano.2c09338
40. Chen, C.-L. and C.-L. Dong, *Characterization of the Electronic Structure of Spinel Superconductor LiTi2O4 using Synchrotron X-ray.* Superconductors: New Developments, 2015: p. 17. http://dx.doi.org/10.5772/58655
41. He, G., et al., *Anisotropic electron-phonon coupling in the spinel oxide superconductor $LiTi_2O_4$.* Physical Review B, 2017. **95**(5): p. 054510. https://doi.org/10.1103/PhysRevB.95.054510
42. Chen, D., et al., *Raman study of electron-phonon coupling in thin films of the spinel oxide superconductor $LiTi_2O_4$.* Physical Review B, 2017. **96**(9): p. 094501. https://doi.org/10.1103/PhysRevB.96.094501
43. Moessner, R. and A.P. Ramirez, *Geometrical frustration.* Physics Today, 2006. **59**(2): p. 24-29. https://doi.org/10.1063/1.2186278
44. Nocera, D.G., et al., *Spin frustration in 2D kagome lattices: A problem for inorganic synthetic chemistry.* Chemistry–A European Journal, 2004. **10**(16): p. 3850-3859. https://doi.org/10.1002/chem.200306074
45. Minervini, L., R.W. Grimes, and K.E. Sickafus, *Disorder in pyrochlore oxides.* Journal of the American Ceramic Society, 2000. **83**(8): p. 1873-1878. https://doi.org/10.1111/j.1151-2916.2000.tb01484.x
46. Moshopoulou, E.G., *Superconductivity in the spinel compound LiTi2O4.* Journal of the American Ceramic Society, 1999. **82**(12): p. 3317-3320. https://doi.org/10.1111/j.1151-2916.1999.tb02245.x
47. Feng, C., et al., *Synthesis and properties of Li–Ti–O spinel (LiTi2O4).* Journal of alloys and compounds, 2009. **478**(1-2): p. 767-770. https://doi.org/10.1016/j.jallcom.2008.12.002
48. Sun, C.P., et al., *Magnetic field dependence of low-temperature specific heat of the spinel oxide superconductor ${\mathrm{LiTi}}_{2}{\mathrm{O}}_{4}$.* Physical Review B, 2004. **70**(5): p. 054519. https://doi.org/10.1103/PhysRevB.70.054519657
49. Chopdekar, R.V., et al., *Growth and characterization of superconducting spinel oxide LiTi2O4 thin films.* Physica C: Superconductivity, 2009. **469**(21): p. 1885-1891. https://doi.org/10.1016/j.physc.2009.05.009
50. Okada, Y., et al., *Scanning tunnelling spectroscopy of superconductivity on surfaces of LiTi2O4(111) thin films.* Nature Communications, 2017. **8**(1): p. 15975. https://doi.org/10.1038/ncomms15975
51. Fujisawa, Y., et al., *Visualizing magnetic field-induced rotational electronic symmetry breaking in a spinel oxide superconductor.* arXiv preprint arXiv:2306.06711, 2023. https://doi.org/10.48550/arXiv.2306.06711
52. Fujisawa, Y., et al., *Imaging emergent exotic quasiparticle state in a frustrated transition metal oxide.* arXiv preprint arXiv:2306.06708, 2023. https://doi.org/10.48550/arXiv.2306.06711
53. Jia, Y., et al., *Crystallographic dependent transport properties and oxygen issue in superconducting LiTi2O4 thin films.* arXiv preprint arXiv:1608.06683, 2016. https://doi.org/10.48550/arXiv.1608.06683
54. Dobrovolny, C., L. Laanait, and J. Ruiz, *Surface Transitions of the Semi-Infinite Potts Model II: The Low Bulk Temperature Regime.* Journal of Statistical Physics, 2004. **116**(5): p. 1405-1434. https://doi.org/10.1023/B:JOSS.0000041744.83013.5d
55. Kondo, S., et al., *$LiV_2O_4$: A heavy fermion transition metal oxide.* Physical review letters, 1997. **78**(19): p. 3729. https://doi.org/10.1103/PhysRevLett.78.3729
56. Anisimov, V., et al., *Electronic structure of the heavy fermion metal $LiV_2O_4$.* Physical review letters, 1999. **83**(2): p. 364. https://doi.org/10.1103/PhysRevLett.83.364

57. Grundner, M., et al., *LiV2O4: Hund-Assisted Orbital-Selective Mottness.* arXiv preprint arXiv:2409.17268, 2024. https://doi.org/10.48550/arXiv.2409.17268
58. Hopkinson, J. and P. Coleman, *L i V 2 O 4: Frustration Induced Heavy Fermion Metal.* Physical review letters, 2002. **89**(26): p. 267201. https://doi.org/10.1103/PhysRevLett.89.267201
59. Laad, M., L. Craco, and E. Müller-Hartmann, *Heavy-fermion behavior of the spinel-based transition-metal oxide LiV 2 O 4.* Physical Review B, 2003. **67**(3): p. 033105. https://doi.org/10.1103/PhysRevB.67.033105
60. Miyake, K. and Y. Kuramoto, *Itinerant-localized duality model for heavy fermions and strongly correlated metals.* Physica B: Condensed Matter, 1991. **171**(1-4): p. 20-29. https://doi.org/10.1016/0921-4526(91)90486-X
61. Singh, D., et al., *Electronic structure and heavy-fermion behavior in LiV 2 O 4.* Physical Review B, 1999. **60**(24): p. 16359. https://doi.org/10.1103/PhysRevB.60.16359
62. Johnston, D.C., *Thermodynamics of the nonrelativistic free-electron Fermi gas in one, two, and three dimensions from the degenerate to the nondegenerate temperature regime.* arXiv preprint arXiv:2012.07663, 2020. https://doi.org/10.48550/arXiv.2012.07663
63. Andrade, B. and R. Gammag, *Analytical form of the density of states of a Fermi gas in low dimensions.* arXiv preprint arXiv:2410.24180, 2024. https://doi.org/10.48550/arXiv.2410.24180
64. Kim, W.J., et al., *Geometric frustration of Jahn-Teller order and ice rules in the infinite-layer lattice.* 2022. https://doi.org/10.21203/rs.3.rs-1524274/v1
65. Kondo, S., D. Johnston, and L. Miller, *Synthesis, characterization, and magnetic susceptibility of the heavy-fermion transition-metal oxide LiV 2 O 4.* Physical Review B, 1999. **59**(4): p. 2609. https://doi.org/10.1103/PhysRevB.59.2609
66. Yang, H., et al., *Evidence of a Kondo lattice quantum critical point and of non-Fermi liquid behavior in the intercalated layered system $V_{5}S S_{8}$.* arXiv preprint arXiv:2408.09956, 2024. https://doi.org/10.48550/arXiv.2408.09956
67. Kontani, H., *Generalized Kadowaki–Woods Relation in Heavy Fermion Systems with Orbital Degeneracy.* Journal of the Physical Society of Japan, 2004. **73**(3): p. 515-518. https://doi.org/10.1143/jpsj.73.515
68. Backes, S., et al., *Ab initio study on heavy-fermion behavior in LiV 2 O 4: Role of Hund's coupling and stability.* Physical Review B, 2025. **111**(4): p. L041102. https://doi.org/10.1103/PhysRevB.111.L041102
69. Gong, B.-C., et al., *Degenerate antiferromagnetic states in spinel oxide LiV2O4.* Chinese Physics B, 2020. **29**(7): p. 077508. https://doi.org/ 10.1088/1674-1056/ab9617
70. Okabe, H., et al., *Metallic spin-liquid-like behavior of LiV 2 O 4.* Physical Review B, 2019. **99**(4): p. 041113. https://doi.org/10.1103/PhysRevB.99.041113
71. Lee, S.H., et al., *Spin Fluctuations in a Magnetically Frustrated Metal ${\mathrm{LiV}}_{2}{O}_{4}$.* Physical Review Letters, 2001. **86**(24): p. 5554-5557. https://doi.org/10.1103/PhysRevLett.86.5554
72. Krimmel, A., *Electronically highly correlated ternary transition metal oxides*. 2005: diplom. de. https://books.google.co.in/books?id=1TNqAQAAQBAJ&dq=Krimmel,+A.,+Electronically+highly+correlated+ternary+transition+metal+oxides.+2005:+diplom.+de.&lr=&source=gbs_navlinks_s
73. Kadono, R., et al., *Quasi-One-Dimensional Spin Dynamics in LiV2O4: One-to-Three-Dimensional Crossover as a Possible Origin of Heavy Fermion State.* journal of the physical society of japan, 2011. **81**(1): p. 014709. https://doi.org/10.1143/JPSJ.81.014709
74. Okamoto, Y., et al., *Band Jahn-Teller Instability and Formation of Valence Bond Solid<? format?> in a Mixed-Valent Spinel Oxide LiRh 2 O 4.* Physical review letters, 2008. **101**(8): p. 086404. https://doi.org/10.1103/PhysRevLett.101.086404

75. Mokdad, J., et al., *Structural, magnetic, and insulator-to-metal transitions under pressure in the Ga V 4 S 8 Mott insulator: A rich phase diagram up to 14.7 GPa.* Physical Review B, 2019. **100**(24): p. 245101. https://doi.org/10.1103/PhysRevB.100.245101
76. Khomskii, D., *Orbital physics: Glorious past, bright future.* ECS Journal of Solid State Science and Technology, 2022. **11**(5): p. 054004. https://doi.org/ 10.1149/2162-8777/ac6906
77. Piekarz, P., A. Oleś, and K. Parlinski, *Comparative study of the electronic structures of Fe3O4 and Fe2SiO4.* arXiv preprint arXiv:1007.2340, 2010. https://doi.org/10.48550/arXiv.1007.2340
78. Talanov, V., et al., *Theory of structural phase transition in MgTi 2 O 4.* Crystallography Reports, 2015. **60**: p. 101-110. https://doi.org/10.1134/S1063774515010253
79. Yang, L., et al., *Two-orbital degeneracy lifted local precursor to a metal-insulator transition in MgTi 2 O 4.* Physical Review B, 2020. **102**(23): p. 235128. https://doi.org/10.1103/PhysRevB.102.235128
80. Zhu, Y., et al., *An Electron Paramagnetic Resonance Study of the Orbital Ordering Compound MgTi 2 O 4.* Applied Magnetic Resonance, 2015. **46**: p. 505-513. https://doi.org/10.1007/s00723-015-0653-8
81. Leoni, S., et al., *Orbital-spin order and the origin of structural distortion in MgTi 2 O 4.* Physical Review B—Condensed Matter and Materials Physics, 2008. **78**(12): p. 125105. https://doi.org/10.1103/PhysRevB.78.125105
82. Shorikov, A., et al., *Orbital-selective pressure-driven metal to insulator transition in FeO from dynamical mean-field theory.* Physical Review B—Condensed Matter and Materials Physics, 2010. **82**(19): p. 195101. https://doi.org/10.1103/PhysRevB.82.195101
83. Chen, J., et al., *Photoinduced insulator-metal transition in paramagnetic (V 1− x Cr x ) 2 O 3.* Physical Review B, 2024. **110**(4): p. 045117. https://doi.org/10.1103/PhysRevB.110.045117
84. Tomczak, J.M. and S. Biermann, *Optical properties of correlated materials: Generalized Peierls approach and its application to VO 2.* Physical Review B—Condensed Matter and Materials Physics, 2009. **80**(8): p. 085117. https://doi.org/10.1103/PhysRevB.80.085117.
85. Croft, M., et al., *Universality in one-dimensional orbital wave ordering in spinel and related compounds: an experimental perspective.* New Journal of Physics, 2007. **9**(4): p. 86. https://doi.org/ 10.1088/1367-2630/9/4/086
86. Rivas-Murias, B., et al., *Rapidly fluctuating orbital occupancy above the orbital ordering transition in spin-gap compounds.* Physical Review B, 2011. **83**(16): p. 165131. https://doi.org/10.1103/PhysRevB.83.165131
87. Di Matteo, S., G. Jackeli, and N.B. Perkins, *Valence-bond crystal and lattice distortions in a pyrochlore antiferromagnet with orbital degeneracy.* Physical Review B, 2005. **72**(2): p. 024431. https://doi.org/10.1103/PhysRevB.72.024431
88. Yang, L., et al., *Two-orbital degeneracy lifted local precursor to a metal-insulator transition in ${\mathrm{MgTi}}_{2}{\mathrm{O}}_{4}$.* Physical Review B, 2020. **102**(23): p. 235128. https://doi.org/10.1103/PhysRevB.102.235128
89. Hotta, T., *Orbital ordering phenomena in d-and f-electron systems.* Reports on Progress in Physics, 2006. **69**(7): p. 2061. https://doi.org/ 10.1088/0034-4885/69/7/R02
90. Li, Q., et al., *Evolution of orbital excitations from insulating to superconducting ${\mathrm{MgTi}}_{2}{\mathrm{O}}_{4}$ films.* Physical Review B, 2023. **107**(12): p. L121108. https://doi.org/10.1103/PhysRevB.107.L121108
91. Rahaman, A., et al., *Surface-phase superconductivity in a Mg-deficient V-doped ${\mathrm{MgTi}}_{2}{\mathrm{O}}_{4}$ spinel.* Physical Review B, 2023. **107**(24): p. 245124. https://doi.org/10.1103/PhysRevB.107.245124
92. Leonov, I., et al., *Charge order and spin-singlet pair formation inTi4O7.* Journal of Physics: Condensed Matter, 2006. **18**(48): p. 10955. https://doi.org/10.1088/0953-8984/18/48/022
93. Hikihara, T. and Y. Motome, *Orbital and spin interplay in spin-gap formation in pyroxene A Ti Si 2 O 6 (A= Na, Li).* Physical Review B—Condensed Matter and Materials Physics, 2004. **70**(21): p. 214404. https://doi.org/10.1103/PhysRevB.70.214404

94. Васильев, А., М. Маркина, and Е. Попова, *Спиновая щель в низкоразмерных магнетиках (Обзор).* Физика низких температур, 2005. Vasiliev, A. N., M. M. Markina, and E. AND. Popova. "Spin slit in low-sized magnets (Review)." Low temperature physics (2005). http://dspace.nbuv.gov.ua/handle/123456789/121749
95. Li, H., R. Clay, and S. Mazumdar, *The paired-electron crystal in the two-dimensional frustrated quarter-filled band.* Journal of Physics: Condensed Matter, 2010. **22**(27): p. 272201. https://doi.org/ 10.1088/0953-8984/22/27/272201
96. Zhang, Z., et al., *Spinel oxide enables high-temperature self-lubrication in superalloys.* Nature Communications, 2024. **15**(1): p. 10039. https://doi.org/10.1038/s41467-024-54482-w
97. Wang, M., et al., *Spinel-Type Metal Oxides with Tailored Amorphous/Crystalline Heterointerfaces for Enhanced Electrocatalytic Water Splitting.* Advanced Functional Materials, 2024. **34**(51): p. 2410439. https://doi.org/10.1002/adfm.202410439
98. Elbaz, Y. and M. Caspary Toroker, *From density functional theory to machine learning predictive models for electrical properties of spinel oxides.* Scientific Reports, 2024. **14**(1): p. 12150. https://doi.org/10.1038/s41598-024-62788-4
99. Sun, Z., et al., *High entropy spinel-structure oxide for electrochemical application.* Chemical Engineering Journal, 2022. **431**: p. 133448. https://doi.org/10.1016/j.cej.2021.133448
100. Li, A., et al., *Enhancing the stability of cobalt spinel oxide towards sustainable oxygen evolution in acid.* Nature Catalysis, 2022. **5**(2): p. 109-118. https://doi.org/10.1038/s41929-021-00732-9
101. Rajesh, V.C., et al., *Growth and characterization of superconducting spinel oxide LiTi2O4 thin films.* Physica C: Superconductivity, 2009. **469**(21): p. 1885-1891. https://doi.org/10.1016/j.physc.2009.05.009
102. Jin, K., et al., *Anomalous magnetoresistance in the spinel superconductor LiTi2O4.* Nature Communications, 2015. **6**(1): p. 7183. https://doi.org/10.1038/ncomms8183
103. Jiang, T., et al., *Multiphysics simulations of lithiation-induced stress in Li1+ x Ti2O4 electrode particles.* The Journal of Physical Chemistry C, 2016. **120**(49): p. 27871-27881. https://doi.org/10.1021/acs.jpcc.6b09775
104. Thauer, E., et al., *Sol-gel synthesis of Li3VO4/C composites as anode materials for lithium-ion batteries.* Journal of Alloys and Compounds, 2021. **853**: p. 157364. https://doi.org/10.1016/j.jallcom.2020.157364
105. Dey, D., T. Maitra, and A. Taraphder, *Possible routes to superconductivity in the surface layers of V-doped ${\mathrm{Mg}}_{1\ensuremath{-}\ensuremath{\delta}}{\mathrm{Ti}}_{2}{\mathrm{O}}_{4}$ through multiple charge transfers and suppression of Jahn-Teller activity.* Physical Review B, 2023. **107**(17): p. 174515. https://doi.org/10.1103/PhysRevB.107.174515
106. Zhao, Q., et al., *Spinels: controlled preparation, oxygen reduction/evolution reaction application, and beyond.* Chemical reviews, 2017. **117**(15): p. 10121-10211. https://doi.org/10.1021/acs.chemrev.7b00051
107. Srikala, D., S. Kaushik, and M. Verma, *Overview on Spinel Oxides: Synthesis and Applications in Various Fields.* Physics of the Solid State, 2024. **66**(9): p. 327-340. https://doi.org/10.1134/S1063783424601073
108. Amiri, M., M. Salavati-Niasari, and A. Akbari, *Magnetic nanocarriers: evolution of spinel ferrites for medical applications.* Advances in colloid and interface science, 2019. **265**: p. 29-44. https://doi.org/10.1016/j.cis.2019.01.003
109. Daptary, G.N., et al., *Effect of spin-orbit interaction on the vortex dynamics in LaAlO 3/SrTiO 3 interfaces near the superconducting transition.* Physical Review B, 2019. **100**(12): p. 125117. https://doi.org/10.1103/PhysRevB.100.125117
110. Daptary, G.N., et al., *Effect of multiband transport on charge carrier density fluctuations at the LaAlO 3/SrTiO 3 interface.* Physical Review B, 2018. **98**(3): p. 035433. https://doi.org/10.1103/PhysRevB.98.035433

111. Daptary, G.N., et al., *Effect of microstructure on the electronic transport properties of epitaxial CaRuO3 thin films.* Physica B: Condensed Matter, 2017. **511**: p. 74-79. https://doi.org/10.1016/j.physb.2017.02.005
112. Daptary, G.N., et al., *Correlated non-Gaussian phase fluctuations in LaAlO 3/SrTiO 3 heterointerfaces.* Physical Review B, 2016. **94**(8): p. 085104. https://doi.org/10.1103/PhysRevB.94.085104
113. Daptary, G.N., et al., *Probing a spin-glass state in SrRuO 3 thin films through higher-order statistics of resistance fluctuations.* Physical Review B, 2014. **90**(11): p. 115153. https://doi.org/10.1103/PhysRevB.90.115153
114. Daptary, G.N., et al., *Conductivity noise across temperature-driven transitions of rare-earth nickelate heterostructures.* Physical Review B, 2019. **100**(12): p. 125105. https://doi.org/10.1103/PhysRevB.100.125105